# Coupling geologically consistent geostatistical history matching with parameter uncertainty quantification[1]

by E. Barrela[2], V. Demyanov[3], L. Azevedo[2]

[2]CERENA – Centro de Recursos Naturais e Ambiente, DECivil, Instituto Superior Técnico, Lisbon University, Av. Rovisco Pais, 1049-001 Lisbon, Portugal. e-mail: eduardo.barrela@tecnico.ulisboa.pt, leonardo.azevedo@tecnico.ulisboa.pt

[3]Institute of Petroleum Engineering, Heriot-Watt University, Riccarton, EH144AS, UK: v.demyanov@hw.ac.uk

Corresponding Author:

Eduardo Barrela

CERENA – Centro de Recursos Naturais e Ambiente, Pavilhão de Minas

Instituto Superior Técnico

Av. Rovisco Pais, 1049-001 Lisboa

Portugal

Phone: +351218417835

eduardo.barrela@tecnico.ulisboa.pt

**Abstract**

Iterative geostatistical history matching uses stochastic sequential simulation to generate and perturb subsurface Earth models to match historical production data. The areas of influence around each well are one of the key factors in assimilating model perturbation at each iteration. The resulting petrophysical model properties are conditioned to well data with respect to large-scale geological parameters such as spatial continuity patterns and their probability distribution functions. The objective of this work is twofold: (i) to identify geological and fluid flow consistent areas of influence for geostatistical assimilation; and (ii) to infer large-scale geological uncertainty along with the uncertainty in the reservoir engineering parameters through history matching. The proposed method is applied to the semi-synthetic Watt field. The results show better match of the historical production data using the proposed regionalization approach when compared against a standard geometric regionalization approach. Tuning large-scale geological and engineering parameters, as represented by variogram ranges, property distributions and fault transmissibilities, improves the production match and provides an assessment over the uncertainty and impact of each parameter in the production of the field.

# 1   Introduction

History matching is the process of calibrating a numerical three-dimensional reservoir model through perturbation on the geological (e.g., the spatial distribution of the subsurface petrophysical properties of interest) and dynamic reservoir parameters (e.g., fault transmissibility, relative permeability) based on the mismatch between the production response given by the model and the observed historical production (Oliver & Chen 2011).

History matching is an inverse problem with multiple possible solutions, where the relationships between the reservoir model parameters and the resulting production data are complex and highly non-linear (Oliver et al. 2008). Due to the ill-posed nature of history matching there are multiple similarly good solutions, consisting of several different subsurface models originating the same production response and that may match considerably well the observed production data (Carrera et al. 2005). Consequently, many combinations of model parameters can provide equally good history matching results while being located in different regions of the model parameter space (Vincent et al. 1999). The challenge of history matching is to identify these multiple minima, which becomes difficult due to the non-uniqueness of the reservoir model parameters and their large number (Arnold et al. 2013). Reliable uncertainty quantification is based on finding those solutions by effective model perturbation (manual or automatic). Automatic history matching methodologies can be subdivided into three different categories: data assimilation, stochastic optimization and perturbation methods.

Common examples of data assimilation methods are the Ensemble Kalman filtering (Evensen et al. 2007) and ensemble smoothing (Emerick & Reynolds 2013), where the prior probability distribution of the model parameters is updated sequentially in time, by means of ensemble perturbation according to the experimental covariance matrices computed between observed and simulated data. The resulting optimal model ensures that the first and second order statistical moments (mean and covariance) of the posterior distribution and the dynamic states are conditioned to the available data. Ensemble methods are computationally efficient in reaching fast convergence but bear a risk of underestimation of posterior variance.

Stochastic optimization methods are designed to explore the model parameter space to find multiple good-fitting models and are among the most common methods in automatic history matching frameworks. These methods are good at exploring the space of possible solutions in respect to the parameter combinations and accurate Bayesian inference. Some examples of stochastic optimization algorithm application in history matching workflows are the Genetic Algorithm (Ballester & Carter 2007; Erbas & Christie 2007), Evolutionary Strategies (Schulze-Riegert et al. 2001), Particle Swarm Optimization (PSO) (Kathrada 2009; Mohamed et al. 2009), Scatter Search (de Sousa 2007) and Differential Evolution (Hajizadeh et al. 2011). Other examples of commonly used stochastic sampling algorithms for uncertainty quantification are the Hamiltonian Monte Carlo algorithm (Duane et al. 1987), the Neighborhood Algorithm (Sambridge 1999) and the Multi–Objective PSO algorithm, which is an extension of the PSO algorithm for the case of multi–objective optimization (Hajizadeh et al. 2011).

Perturbation methods include the subset of methodologies aimed at finding an optimal solution by applying global or local perturbation of the models, relying on a deformation parameter. Examples of perturbation methodologies applied to history matching workflows are Gradual Deformation (Hu et al. 2001; Roggero & Hu 1998), probability perturbation (Caers & Hoffman 2006) and stochastic sequential simulation and co-simulation (Barrela et al. 2018; Caeiro et al. 2015; Carneiro et al. 2018; Le Ravalec-Dupin & Da Veiga 2011; Mata-Lima 2008).

Such methods are particularly promising considering geologically consistent history matching workflows. Within this framework, geostatistical modelling presents itself as a tool suited for such methodologies since it provides a way of integrating, into the history matching workflow, essential domain knowledge like differently scaled data and geological knowledge, by means of auxiliary variables (Cosentino 2001). Combined with the information provided by production data, geostatistical modelling allows the development of techniques capable of addressing perturbation on the various geological scales: the spatial distribution of the geological units, definition of the structural model and the spatial distribution of the petrophysical e properties. With these methods the resulting history matched models reproduce the existing well data, histograms and the spatial distribution, as revealed by a variogram model inferred from available direct measurements for the petrophysical properties of interest.

Several methods have been focusing on integrating geological consistency within automatic history matching workflows (Demyanov et al. 2018; Oliver & Chen 2011). Machine learning approaches have also been recently used to address the problem of geological realism in history matching (e.g., Demyanov et al. 2008, 2012, 2015).

The hierarchical nature of uncertainty, when the inferred parameter uncertainty needs to be decoupled between geologically interpretable scales, is one of the remaining challenges in geologically consistent history matching. A hierarchical approach in describing uncertainty is illustrated in a benchmark history matching data set in Arnold et al. (2013). The multi-scale challenge can be illustrated considering multiple equally probable geostatistical realizations as a description of uncertainty. The range of geostatistical realizations handles the small-scale uncertainty only regarded as the heterogeneous non-uniqueness, reproducing the first and second order statistics, along with the spatial continuity pattern as described by a variogram model. This assumes that there is no uncertainty related to these parameters, which is hardly true in real case applications. Wells are often located preferentially along pay zones, biasing the available petrophysical property statistics. In addition, the limited number of existing wells does not allow the computation of reliable horizontal experimental variograms and, therefore, the resulting models are plagued with uncertainties.

We introduce herein an iterative geostatistical history matching (GHM) technique via geologically and dynamically consistent perturbation and assimilation through iterations. The technique is coupled with adaptive stochastic sampling for the simultaneous uncertainty quantification of large-scale geological and engineering properties that are conventionally considered as uncertainty free in conventional GHM techniques. This method is applied to a semi–synthetic case study based on a braided–river depositional environment (Watt field) (Arnold et al. 2013).

## 2  Methodology

By encompassing a two–stage optimization process, the proposed GHM technique addresses simultaneously different scales of uncertainty based on the following key ideas: (i) model regionalization based on geological and dynamic criteria; (ii) integration of stochastic adaptive sampling within the iterative procedure, to assess uncertain large-scale geological and dynamic engineering parameters.

The first stage (**Fig. 1**), includes generation of multiple porosity ($\Phi$) and permeability (K) realizations using direct sequential simulation with multi-local distribution functions and (Nunes et al. 2016) and direct sequential co-simulation with joint probability distributions (Horta & Soares 2010). At the end of each iteration, a patchwork model for each property is built by selecting the pairs of Φ and K realizations that ensure the minimum misfit in terms of production responses for a given well area of influence. These patchwork models are updated iteratively and used as secondary variables in the co-simulation of Φ and K for the next iteration. The second stage, encompasses the first loop and is responsible for exploring the uncertainty in the large-scale geological parameter space: parameters of the variogram model, Φ and K global distributions (i.e., histograms) and fault transmissibilities that interconnect the different regions of the reservoir. These uncertain parameters are sampled from prior distribution using PSO. At each iteration, the PSO algorithm optimizes the uncertain large-scale parameters, based on the mismatch between historical and simulated production variables. The resulting parameters are then used as input for the first stage of the method to proceed with the computation and assimilation between multiple equally probable geostatistical realizations (**Fig. 2**). The process iterates until a final number of iterations is reached, or when convergence of the parameter perturbation and misfit value is above a pre–defined threshold.

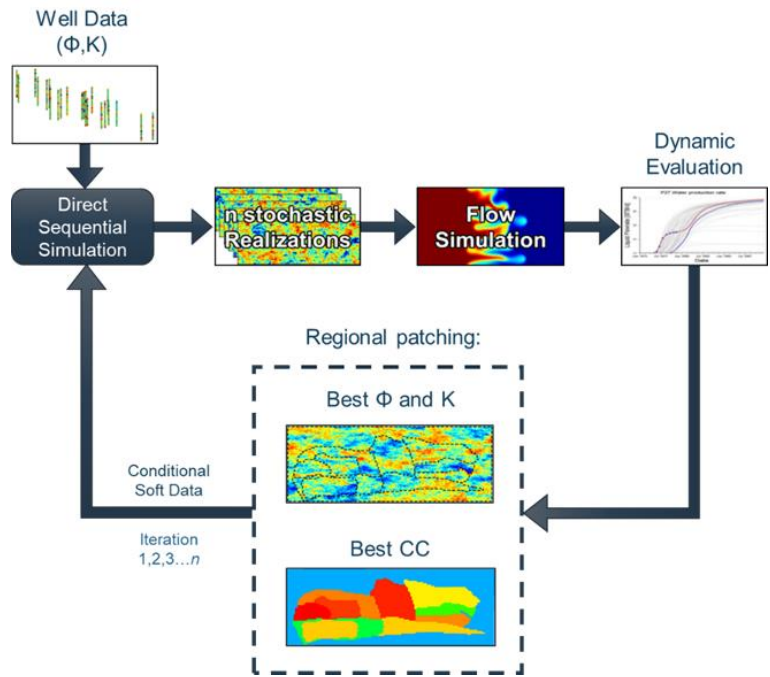

**Fig. 1.** Schematic representation of the first stage of the proposed multi-scale GHM method

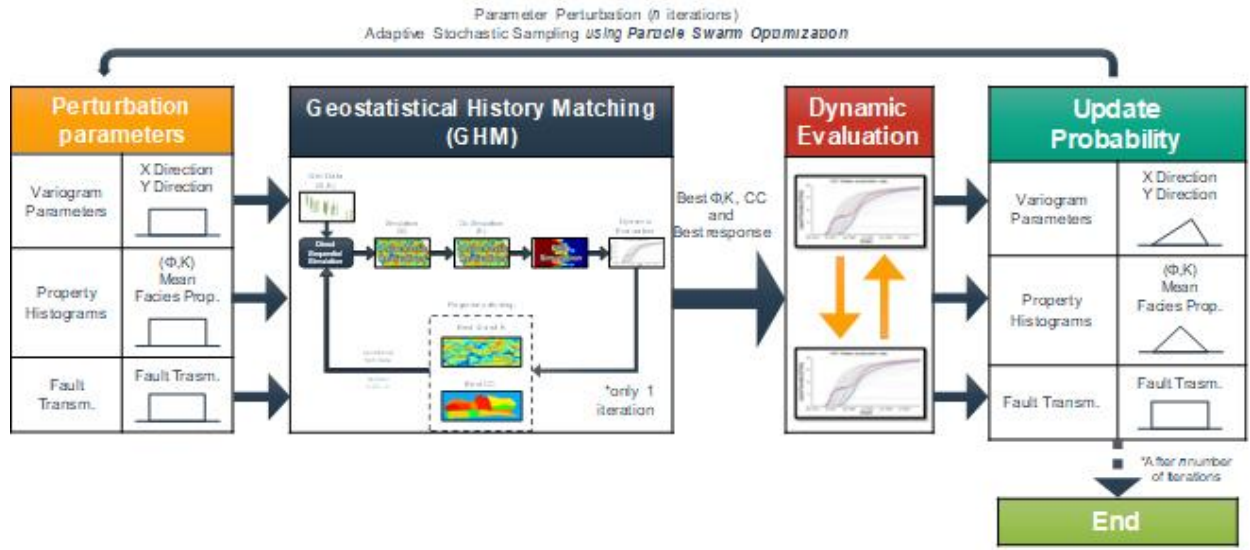

**Fig. 2.** Coupling of adaptive stochastic sampling algorithm with the GHM technique shown in detail in **Fig. 1**

## 2.1 Geological and dynamic consistent iterative geostatistical history matching

Previous works on regionalization–based history matching algorithms done by Caers & Hoffman (2006), Mata–Lima (2008) and Le Ravalec-Dupin & Da Veiga (2011), introduced methods of property perturbation and assimilation using geostatistical simulation. The present work investigates the impact of regionalization (i.e., the definition of areas of influence per well) in GHM and proposes the introduction of geological and dynamic consistent regionalization. This allows the local perturbation of the geological trends of the reservoir, in terms of Φ and K, under a geological and dynamic regionalization model perturbed by stochastic sequential co-simulation.

In GHM, at the end of each iteration, the new ensemble of Φ and K models is generated through stochastic sequential co-simulation using a patchwork model of Φ and K as secondary variables. These models are built by selecting the pairs of realizations that, for a given iteration, ensure the minimum misfit in terms of simulated production data for each area of influence (e.g., polygonal/radial areas of influence) around the existing wells. The way this regionalization is performed is one of the main handicaps to these methods, since is often based exclusively on geometric criteria and decoupled from the expected subsurface geology.

Therefore, non-geometric and geologically consistent regionalization criteria to be included in iterative GHM workflows is explored. Accounting for geological consistency during the model perturbation allows to screen out the set of solutions that are able to approximate production history but that are not compatible with, or do not reflect, the real subsurface geology. Furthermore, an adequate regionalization criterion might help in mitigating a common problem of most history matching workflows, which is, the difficulty in obtaining simultaneous matches on every producing well in the reservoir (Hoffman & Caers 2003). Since individual well production often influences the production of neighboring wells, a regionalization criteria capable of capturing individual well (or group of wells) drainage areas, will be able to provide better predictions.

We propose a regionalization pattern based simultaneously on fault presence and production streamlines to tackle simultaneously the inclusion of geological consistency and inter-well production influence (i.e., dynamic behavior) in definition of a regionalization model. Geological consistency is integrated by identifying and delineating different model flow units with the geological support given by fault presence. Furthermore, drainage area of influence for each well, or group of wells, into the regionalization criteria is incorporated by analyzing production streamline patterns on the reservoir, to focus the perturbation in locations that are preponderant to fluid production.

To validate the proposed regionalization approach, the results obtained with a GHM technique using the proposed regionalization criteria are compared against the same history matching methodology with a global perturbation and using pure geometric criteria: Voronoi polygons centered at production well locations (Caeiro et al. 2015).

The proposed methodology is summarized by the following sequence of steps and illustrated in (**Fig. 1**):

i) Regionalization of the reservoir area according to a given fault model and dynamic area of influence of the wells resulting in a model with a region being assigned for each well or for a group of wells;

ii) Simulation of a set of $\Phi$ and co-Simulation of a set of K realizations using DSS, honoring the well data, histograms and spatial distributions as revealed by the variogram;

iii) Evaluation of the dynamic responses for each of pair of realizations and calculation of the mismatch between the dynamic response and real production data, using an objective function: The objective function is formulated as a mathematical expression that measures how close a problem solution (i.e., simulated data) is, towards an optimal value (i.e., observed data). The definition of such a metric is critical to achieve convergence in the iterative procedure. The most commonly used objective function for history matching is the least square norm (Eq. (1); Arnold et al. 2013), which calculates a measure of the misfit between the simulated and observed value, $M$, to be minimized during the iterative procedure, according to:

$$M = min \sum_{t=i}^{n} \frac{(R_{t,obs} - R_{t,sim})^2}{2\sigma^2}, \quad (1)$$

where $R_{(t,obs)}$ is the observed, or historic value of a given variable at timestep $t$, $R_{(t,sim)}$ is the simulated value of a given variable at timestep $t$, $\sigma^2$ is the admissible error for the measured variable, assumed to be Gaussian, independent and constant with time.

iv) Calculation of a linear correlation coefficient between each dynamic response and generation of a cube composed of different correlation coefficients per region (the calculation of the correlation coefficient is explained in detail in the following sub-section). Due to its importance in the evolution, and convergence of the iterative procedure, how the correlation coefficient is computed is detailed in the next sub-section;

v) Composition of patchwork Φ and K volumes, with each region being populated by the properties corresponding realization with the lowest mismatch, calculated in step iii);

vi) Return to step ii), using co–DSS and the cubes calculated in steps iv) and v) as local correlation coefficient and "soft data" (secondary variable), respectively. The algorithm is expected to run up to a maximum number of iterations, or until a pre–defined mismatch value is reached.

### 2.1.1  Correlation coefficient

The convergence from iteration-to-iteration is ensured through geostatistical soft conditioning (i.e., stochastic sequential co-simulation using the patchwork models along with local correlation coefficients as soft data for the generation of a new ensemble of petrophysical models in the next iteration). The calculation of the correlation coefficient is an important aspect of the proposed GHM workflow as it serves as a gauging parameter that controls local variability in terms of property distribution and spatial continuity. Poorly matched regions will be associated with low correlation coefficient values, hence being able to vary more in the following iteration, while regions that produce good match quality, and consequently high correlation coefficients, will only exhibit small-scale variability during the next iteration. By doing patch composition of the best properties and calculating the respective local correlation coefficients, convergence is obtained through iterative generation of subsequent realizations that not only better reproduce the observed production data, but that also assimilate the spatial distribution of unknown model parameters.

For this purpose, the correlation coefficient is calculated based on the misfit value (Eq. (1)). The following sequence of steps describes the calculation of the correlation coefficient being used by the proposed GHM algorithm (**Fig. 3**):

i) For all production time steps $t$ and for the production variable to be matched, calculate the absolute difference $\Delta_{sim}$, between simulated response $R_{(t,sim)}$ and observed data $R_{(t,obs)}$, following:

$$\Delta_{sim} = |R_{(t,obs)} - R_{(t,sim)}|; \qquad (2)$$

ii) We define a threshold value $CC^*$ that establishes a relationship between the correlation coefficient and the accepted error $\sigma_t^2$, and a constant $\sigma_{tol}$, that will define a range $\sigma_{tol} \times \sigma_t^2$, where the correlation coefficient will have values between 0 and 1. The correlation coefficient $CC_t$ is then calculated for each timestep $t$, according to the following:

$$CC_t = \begin{cases} \frac{(CC^*-1)\times\Delta_{sim}}{\sigma_t^2} + 1 & if \quad \Delta_{sim} \leq \sigma_t^2 \\ CC^* \times \left(\frac{\sigma_t^2 - \Delta_{sim}}{(\sigma_{tol}-1)\times\sigma_t^2} + 1\right) & if \quad \sigma_t^2 \leq \Delta_{sim} \leq \sigma_{tol} \times \sigma_t^2; \\ 0 & if \quad \Delta_{sim} \geq \sigma_{tol} \times \sigma_t^2 \end{cases} \qquad (3)$$

iii) For all time steps with available data $t_n$, the final correlation coefficient $CC$ is calculated for each specific variable $var$, according to:

$$CC_{var} = \sum_{i=1}^{n} \left(\frac{t_i}{t_n^2} \times CC_{t_i}\right). \qquad (4)$$

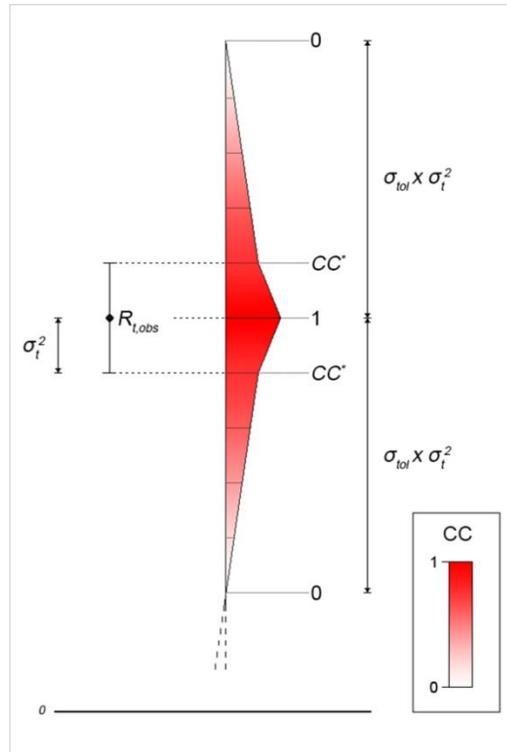

**Fig. 3.** Illustration of the correlation coefficient calculation being used in the proposed GHM workflow

## 2.2 Uncertainty quantification of main geological parameters

By definition, each model of $\Phi$ and K generated during the iterative procedure honors the conditioning well data, the marginal and joint distributions of $\Phi$ and K, which are inferred from the experimental data, and the imposed variogram model that describes the spatial continuity pattern of each property individually. Usually no uncertainty is assumed for these parameters in geostatistical history matching.

To introduce additional levels of uncertainty related to the large-scale geological parameters, adaptive stochastic sampling (i.e., PSO) is coupled within the GHM procedure. This enables uncertainty quantification at two different levels: (i) GHM addresses uncertainty quantification of the local scale petrophysical properties given the spatial continuity pattern and the global distribution; (ii) while adaptive stochastic sampling infers the uncertainty of the large-scale geological parameters used to generate multiple geostatistical realizations (e.g. variogram model, soft conditioning distribution). In addition, adaptive stochastic sampling assesses uncertainty over the reservoir engineering parameters (e.g., fault transmissibilities) that are jointly optimized with the geological parameters.

### 2.2.1 Particle swarm optimization

In the application example shown herein single objective PSO is used as the stochastic adaptive sampling technique, due to its simplicity and the possibility to easily tune a more explorative or exploitative behavior of the algorithm. Developed by Kennedy and Eberhart (1995), PSO is a stochastic optimization technique inspired by the social behavior of bird flocking or fish schooling. The principle behind the PSO algorithm implies considering a search space where a swarm of particles is randomly scattered. The movement of the particles is controlled by the cognitive and social components of their velocity vector, which effectively guide the search to explore the parameter space and home in on the optimal parameter combinations. The particle position and velocity are updated iteratively through the search. The cognitive component ensures every particle retains its previous best position. The social component compares the local and the global best at every step of the algorithm by a fitness function evaluation. From this, a global best position is obtained, which is then responsible (along with all the best positions of all different particles) to converge the solution of the problem to a minimum value.

The particle velocity update for the classical PSO algorithm is achieved using:

$$\hat{V} = w * V + r_1 * nostalgia.(PBest - X) + r_2 * sociality * (GBest - X), \quad (5)$$

where $\hat{V}$ is the updated velocity vector of the particle, $V$ is the old velocity vector of the particle $X$, $PBest$ and $GBest$ are the vectors that respectively describe the current particle position in the search space, their known local best location and their known global best location, $w$ is an inertia coefficient, $r_1$ and $r_2$ are random numbers between 0 and 1 and nostalgia and sociality are external coefficient parameters.

In the present work, PSO is used to perturb large-scale geological parameters as represented by the ranges of the variogram models and the histograms of Φ and K used as conditioning data for the model generation. Fault transmissibility, as an engineering parameters, is also perturbed. The perturbed parameters are presented in the next section (sub-Sect. 3.4).

## 3  Application to the Watt field case study

### 3.1  Dataset description

The Watt field is a semi–synthetic dataset, incorporating synthetic and real data from a North Sea oil field, mimicking a realistic field example seen through appraisal into early development life stages (Arnold et al. 2013). It provides a set of reservoir models representing typical interpretational choices normally found in a standard reservoir geomodeling workflow. The reservoir has a 12.5×2.5 km surface area, elongated along the East–West direction, with a thickness of approximately 190 m, much of which below the OWC, located at 1635 m subsurface. Initial depth of the reservoir is located at 1555 m subsurface with an initial reservoir pressure of 2500 psi. The depositional environment mimics a braided river system, with facies types comprising fluvial channel sands, overbank fine sands, and background shales. The dynamic simulation grid resolution is 226×59×40, with cell size of 100×100×5 m in i-, j- and k- directions respectively.

Well data (Φ and K) is provided for set of 6 appraisal wells. The reservoir is produced under liquid rate control with 16 horizontal production wells located across the central section of the reservoir. Additional 5 horizontal and 2 vertical injector wells are located around the edges of the reservoir to assist production, by providing pressure support (**Fig. 4**).

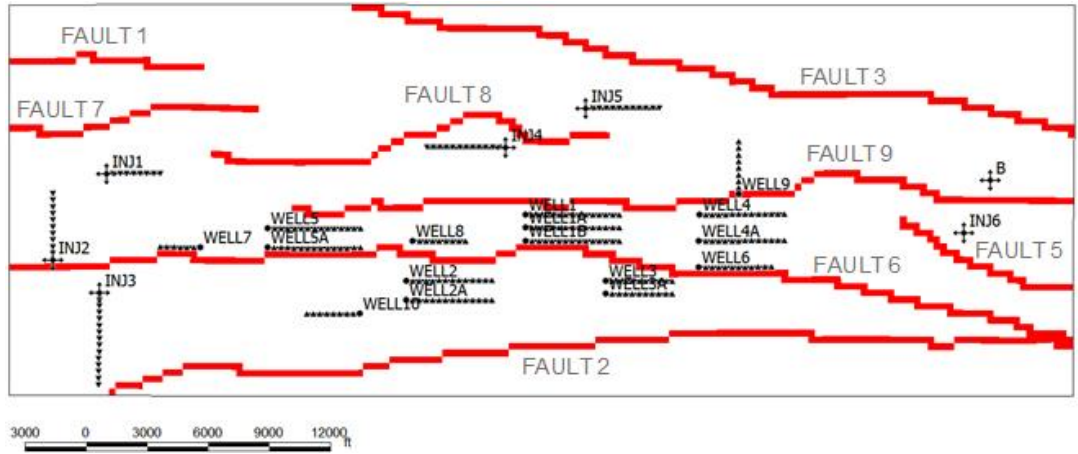

**Fig. 4.** Fault network definition for the Watt field. Fault locations are marked in red, developed well locations are marked in black

Historical production data for the reservoir, generated from an unknown geological scenario, is available for a period of 2903 days (8 years), for all wells, for oil, water and gas rates and bottom-hole pressure (WBHP) to which 15% Gaussian noise was added. **Fig. 5** shows the production history for field oil and water rates, along with bottom-hole pressure for well 1.

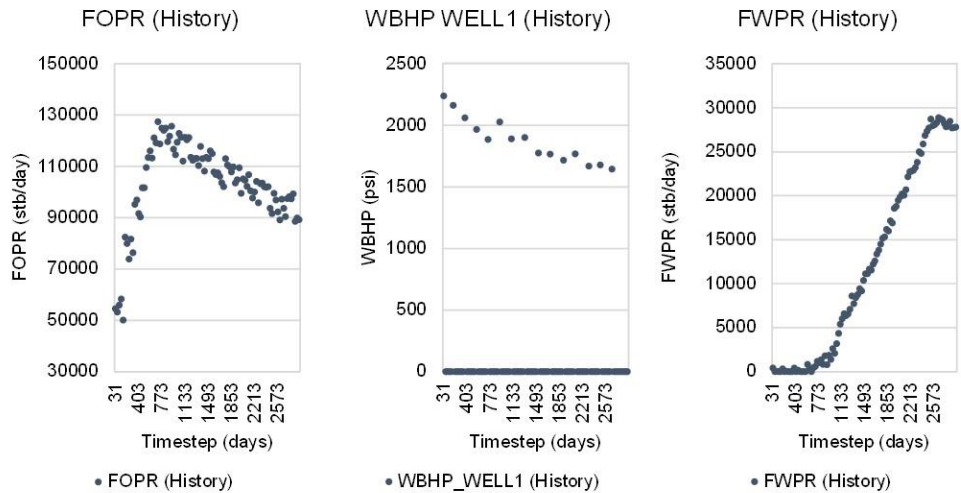

**Fig. 5.** Production history for field oil production rate (FOPR), field water production rate (FWPR) and WBHP for well 1

Out of the ensemble of possible geological scenarios available in the Watt field case, a pair of Φ and K models is selected to be compared against the matched models retrieved by the proposed GHM. Consequently, the production of the reference model does not exactly reproduce the production curves shown in **Fig. 5**, which belong to an unknown subsurface reality. A horizontal section extracted from the reference Φ and K models is shown in **Fig. 6** and **Fig. 7** , while **Fig. 8** provides a top view of the facies model that depicts the reservoir braided-river depositional environment, constraining the spatial distribution of the petrophysical properties of interest into three different geological settings: (i) blue – overbank fine sands; (ii) green – fluvial channel sands; (iii) red – background shales.

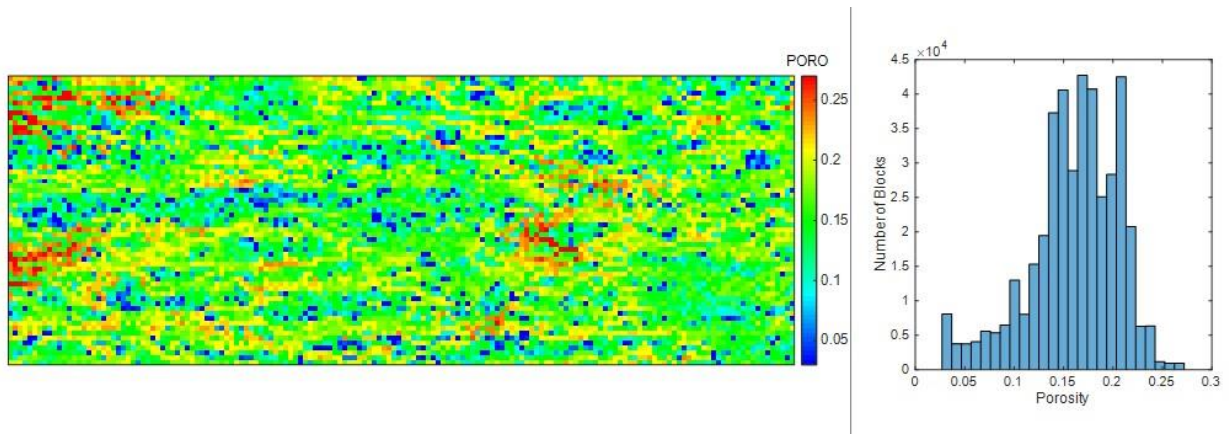

**Fig. 6.** Horizontal section extracted from the Φ field for the reference scenario (left), and corresponding histogram (right)

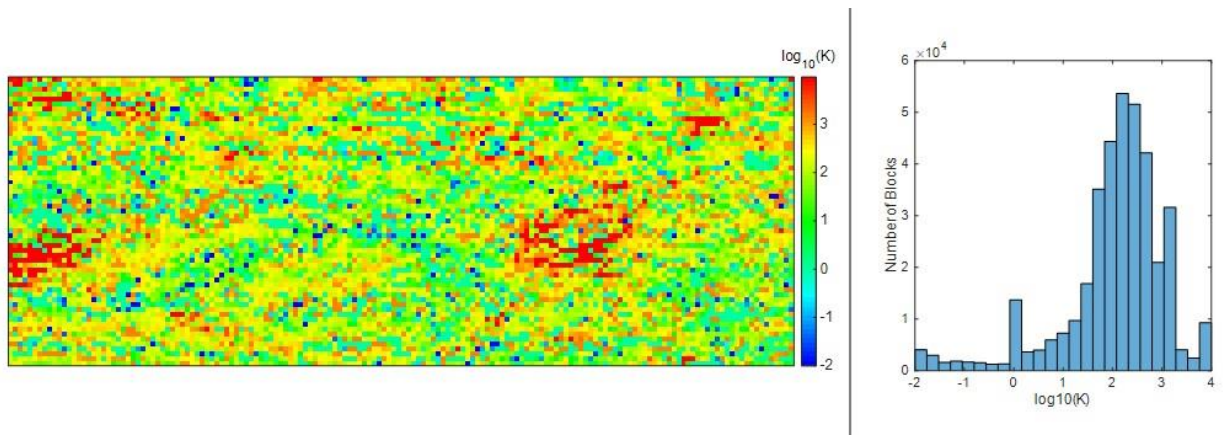

**Fig. 7.** Horizontal section extracted from the K field for the selected base scenario (left), and corresponding histogram for $\log_{10}(k)$ (right)

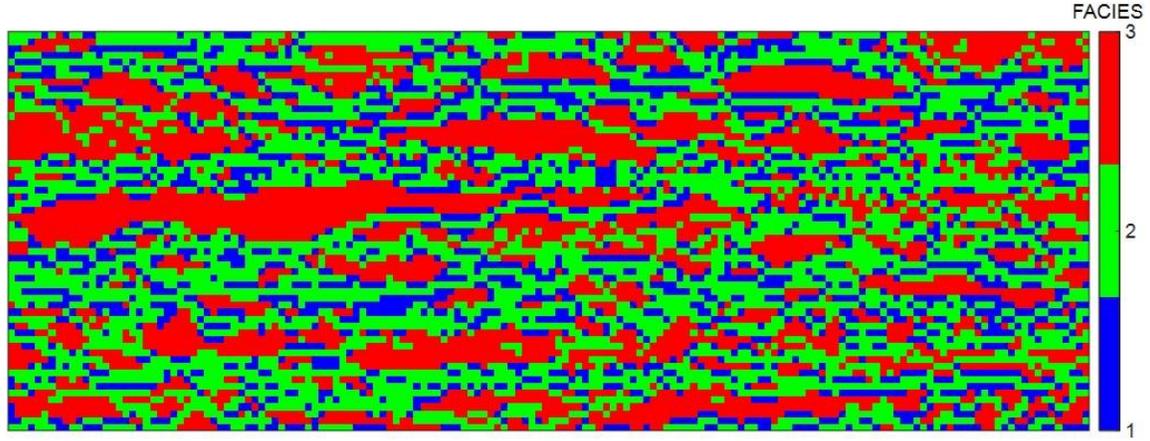

**Fig. 8.** Horizontal section extracted from the saturation regions for the selected base scenario (blue – overbank fine sands, green – fluvial channel sands, red – background shales)

The variables considered for the misfit calculation (M) were field and well rates for oil and water production (FOPR, WOPR, FWPR, WWPR, respectively). The misfits are attributed to each area of influence of individual or group of wells (Eq. 5). In the grid cells unassigned to a well (i.e., non-productive regions), field rates are considered to calculate the misfit for these cells (Eq. 6):

$$M = \sum_{t=i}^{n} \frac{(WOPR_{t,obs} - WOPR_{t,sim})^2}{2\sigma_{WOPR}^2} + \sum_{t=i}^{n} \frac{(WWPR_{t,obs} - WWPR_{t,sim})^2}{2\sigma_{WWPR}^2}, \qquad (6)$$

$$M = \sum_{t=i}^{n} \frac{(FOPR_{t,obs} - FOPR_{t,sim})^2}{2\sigma_{FOPR}^2} + \sum_{t=i}^{n} \frac{(FWPR_{t,obs} - FWPR_{t,sim})^2}{2\sigma_{FWPR}^2}. \qquad (7)$$

### 3.2 Consistent geologic and dynamic regionalization definition

In order to discretize the reservoir model into geologic and dynamic consistent regions, each producing well is assigned to a region delimited by the existing faults in the model. Since fault transmissibility is an unknown and uncertain parameter to be inferred, there might be non-sealing faults that do not exactly reflect the assumed regionalization pattern. Fault regionalization was coupled with the most likely fluid flow pattern of the reservoir using streamline analysis (Kazemi & Stephen 2013; Vargas-Guzmán et al. 2009). As the non-stationary nature of multiphase fluid flow in porous media, implies that streamline paths change over production time, and consequently from model to model, the most important reservoir cells in terms of production are assessed, by running a set of scoping simulations over the available scenarios from the Watt field dataset. Then, the overall streamline paths for the entire production schedules were traced (the same principle could be used in an ensemble of realizations of Φ and K generated with stochastic sequential simulation) (**Fig. 9**).

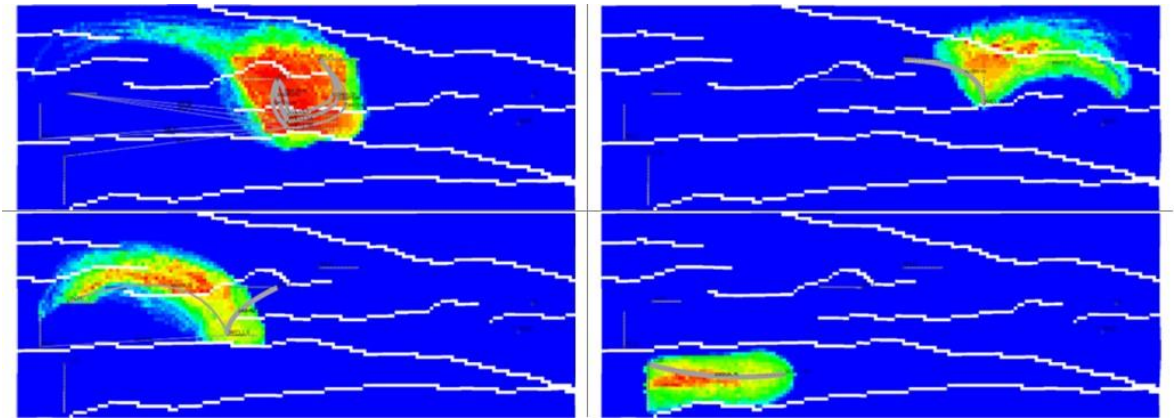

**Fig. 9.** Examples of preferential flow paths (red to blue) for well group 1 (top left), well group 9 (top right), well group 8 (bottom left) and well group 10 (bottom right)

For the case of wells that are closely located and associated in groups, the definition of regions with such criterion is difficult due to high inter-well dependency in terms of production, i.e., there is a strong production trade-off between these wells, from realization to realization. Therefore, wells with high inter-dependency are associated into a single representative region, while also considering fault presence. Together with streamline analysis, the proposed regionalization criteria is obtained, which aims at approximating the drainage areas of each well or groups of wells, simultaneously limited by fault presence and resulting in a set of regions where the geological properties can be updated to match production data with added geologically consistency, while at the same time, giving room for the integration of fault transmissibility (**Fig. 10**).

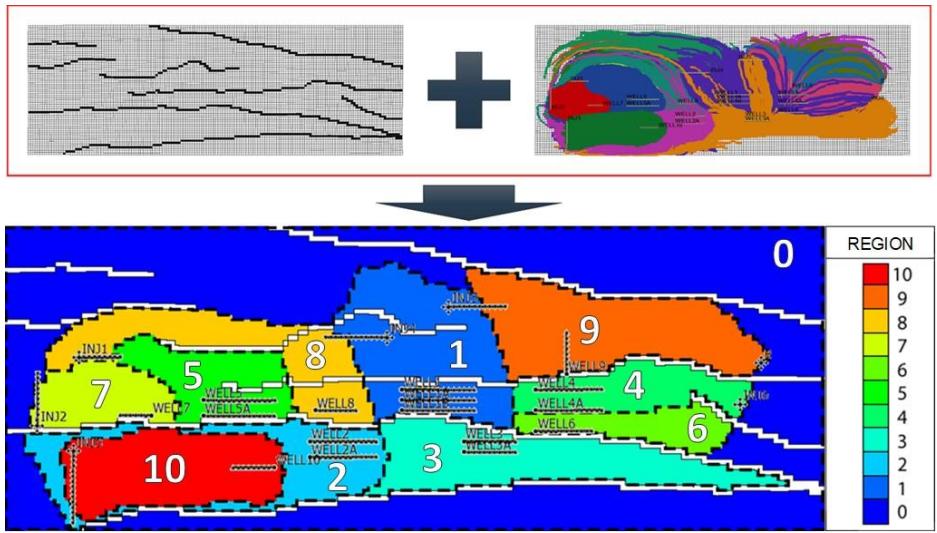

**Fig. 10.** Fault-streamline regionalization criteria using fault network definition and streamline analysis (Top). Geological and dynamic consistent regionalization for each well, or group of wells (Bottom). Non-producing regions are associated to region 0

In the following sub-section, the performance of the proposed geological consistent regionalization methodology is compared against a pure geometric regionalization scenario based on Voronoi polygons centered at the location of the existing wells or group of wells (**Fig. 11**).

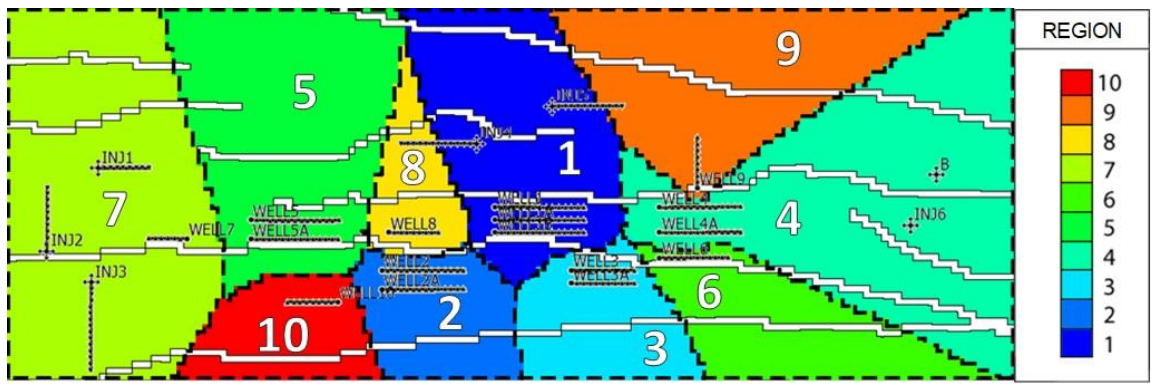

**Fig. 11.** Voronoi regionalization pattern. Individual well or group of well regions are listed from 0 (blue) to 10 (red)

### 3.3 Results: regionalization comparison

The geostatistical history matching run converged in 10 iterations with 50 simulations and co-simulations of Φ and K per iteration. The sum of individual well misfit evolution (Eq. (6)) shows a good convergence in 10 iterations for the average, minimum and standard deviation (**Fig. 12**).

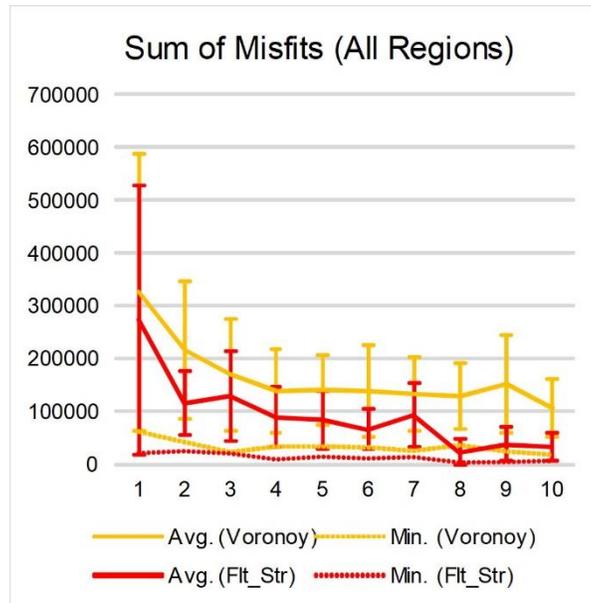

**Fig. 12.** Sum of misfits for all regions. Misfits from Voronoi regionalization are plotted in yellow and fault-streamline in red. Average misfits per iteration are plotted with a continuous line, along with minimum misfit per iteration (dashed line) and misfit standard deviation (error bars)

The results show lower misfits for the fault-streamline regionalization over the entire run, when compared to using Voronoi regionalization, both for field and well variables. The overall trend for the fault-streamline regionalization case features a steeper decrease of average misfit for all wells with the iterations (**Fig. 12**), and a smaller variability when compared against the misfits computed from the Voronoi regionalization. A geological and dynamic regionalization criteria, as represented by the fault-streamline method, is able to lead the algorithm into further exploration of inter-well production scenarios that minimize production mismatch, as illustrated by iterations 7 to 10 on the sum of misfits for all wells (**Fig. 12**). The results from iteration 8 onwards also allow to interpret that the average misfit per iteration for the fault-streamline method is lower than the combined average and standard deviation of the Voronoi regionalization, which supports the conclusion about the better performing capabilities of the proposed regionalization method.

Regarding flow response, **Fig. 13** and **Fig. 14** show the first and last iteration of the results obtained for well oil production by both regionalization methods for wells 2A and 5A.

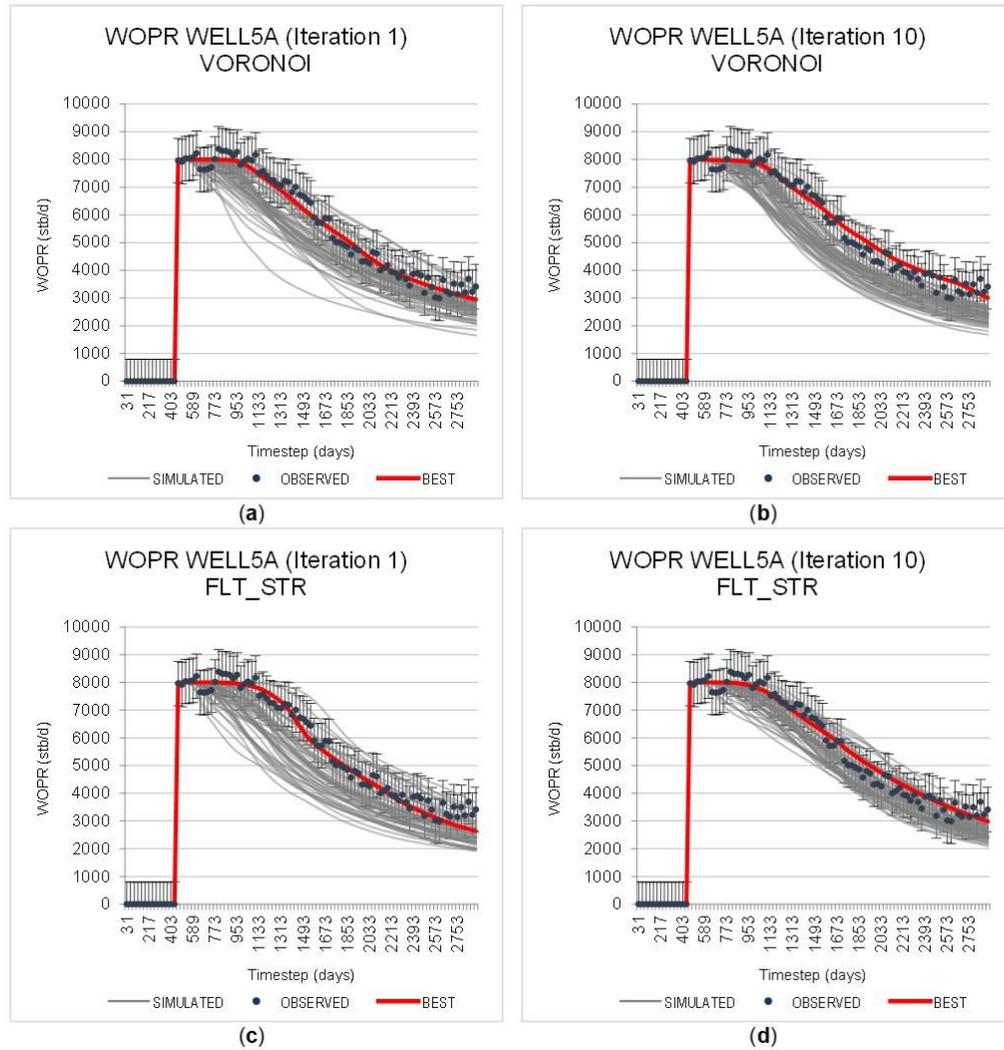

**Fig. 13.** Well oil production for well 5A, using (top) Voronoi and (bottom) fault-streamline regionalization methodologies. First iteration is shown on the left and last iteration on the right. Historical data is plotted in blue dots, along with the admissible error (sigma bars). Best simulation of the iteration is plotted in red and all remaining simulations in grey

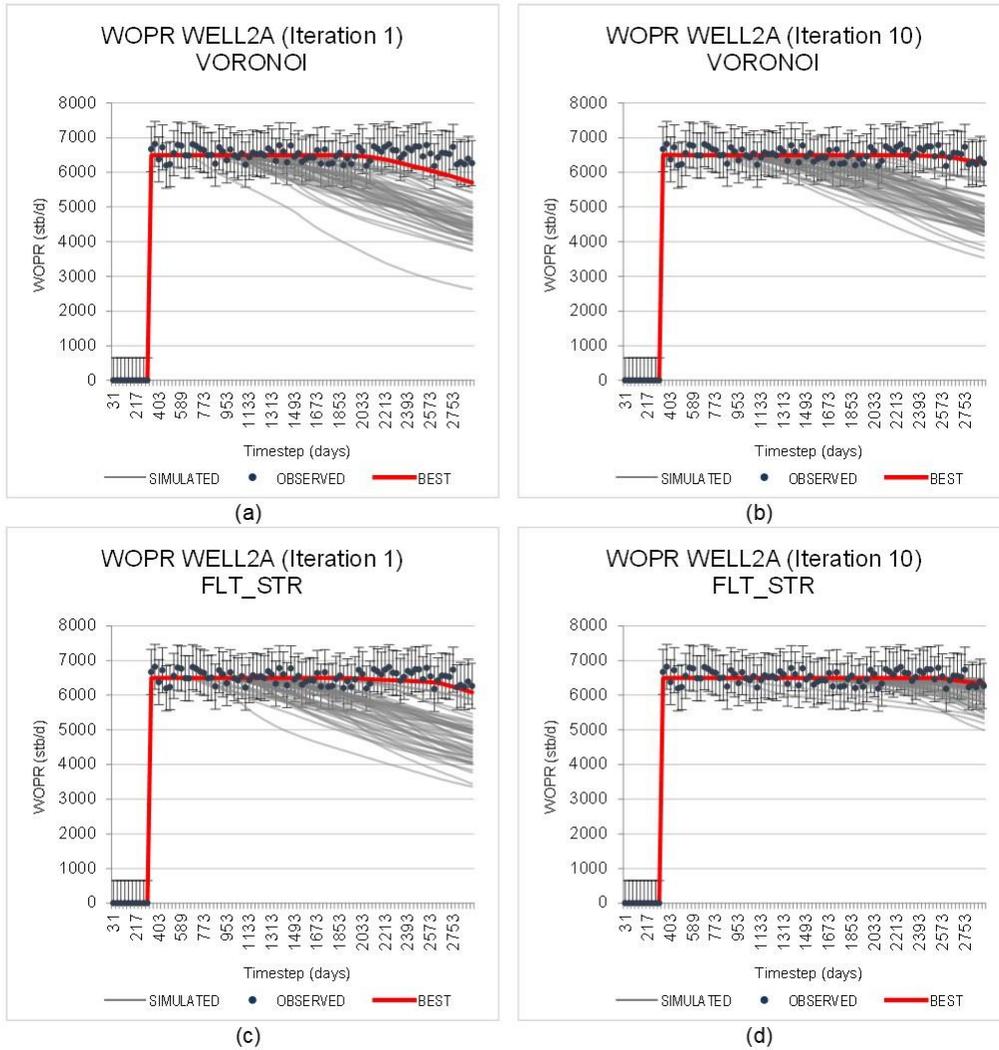

**Fig. 14.** Well oil production for well 2A, using (top) Voronoi and (bottom) fault-streamline regionalization methodologies. First iteration is shown on the left and last iteration on the right. Historical data is plotted in blue dots, along with the admissible error (sigma bars). Best simulation of the iteration is plotted in red and all remaining simulations in grey

At iteration 10, the proposed fault-streamline regionalization methodology produces better results both in terms of best-fit model and by making all realizations fall inside the acceptable range of misfits which is a key aspect for production forecasting and uncertainty quantification. The Voronoi regionalization method fails to converge with the same accuracy (WELL5A, **Fig. 13**) or even diverge (WELL2A, **Fig. 14**).

The same behavior, as interpreted from these two wells, can be interpreted in other regions of the reservoir. **Fig. 15** synthetizes the results in terms of minimum misfit per region for both methods.

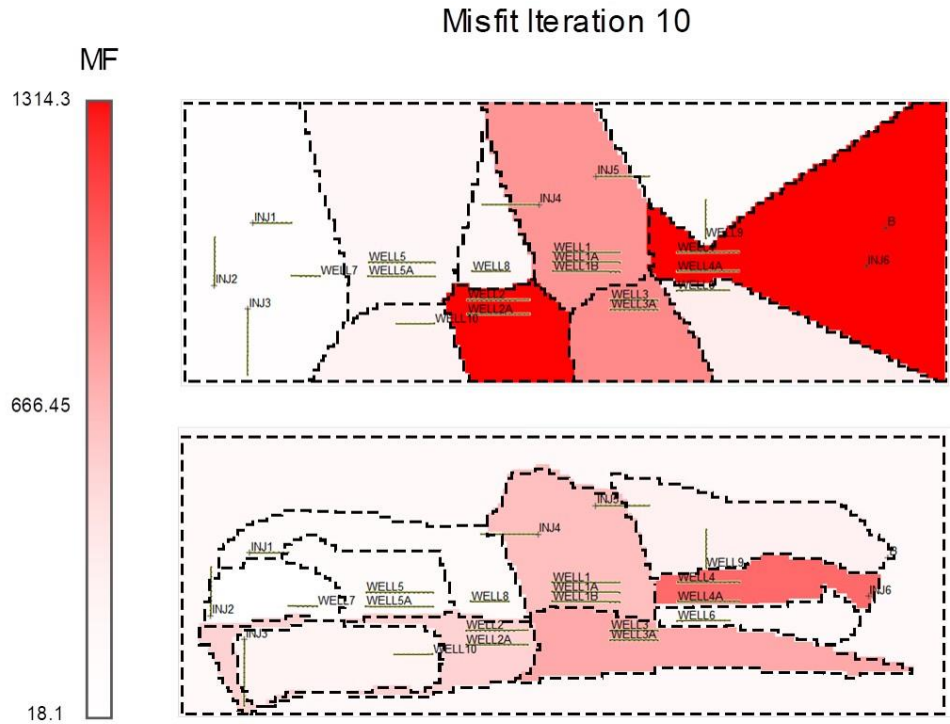

**Fig. 15.** Top field view of the minimum misfits per region achieved at iteration 10

### 3.4   Results: uncertainty assessment of geological and engineering parameters

The results of the previous section show that a geology-based regionalization model allows retrieving more reliable alternative subsurface models capable of matching observed production history and representing a range of plausible geological scenarios that can be used for uncertainty quantification. However, these results were obtained assuming known large-scale geological and engineering parameters as described by a priori Φ and K distributions, variogram models and fault transmissibility.

This section shows the results when coupling adaptive stochastic sampling with iterative GHM as described in Sect. 2.2. At each iteration, five realizations of Φ and K are generated. PSO was selected to optimize the selected parameters over a total of 223 iterations. Within this scope, 17 parameters were perturbed (**Table 1**): horizontal and vertical ranges for the variogram models of Φ (**Fig. 16**) and K, petrophysical property group means and proportions to describe the distribution of both properties (**Fig. 17**); and 7 fault transmissibilities values for faults 2, 3, 5, 6, 7, 8 and 9 (**Fig. 4**).

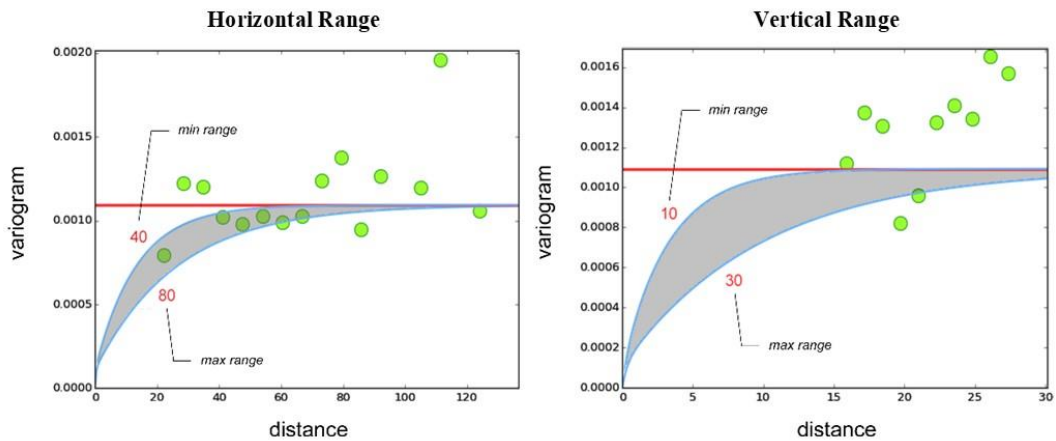

**Fig. 16.** Example of uncertainty boundaries for horizontal (left) and vertical (right) variogram ranges for Φ

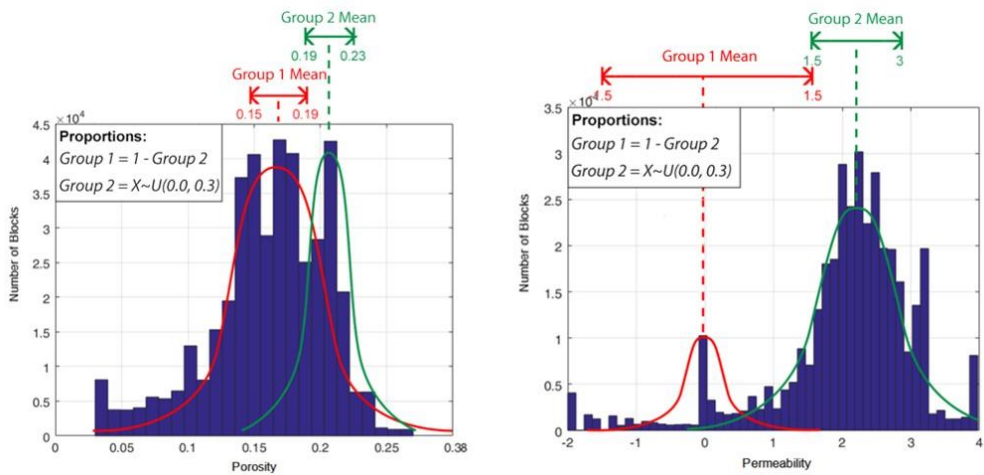

**Fig. 17.** Property group means and proportions ranges (left- Φ, right- K)

Based on the geological beliefs about the reservoir in study, a prior distribution was assigned to each uncertain geological and dynamic parameter (**Table 1**).

| Property | Parameter | Variable name | Distribution type | Prior range |
|---|---|---|---|---|
| **Fault transm.** | All fault transm. | ftrans (2,3,5,6,7,8,9) | Uniform | [0.0, 1.0] |
| **K** | Horizontal range | perm range 1 | Discrete uniform | [40, 80] |
| | Vertical range | perm range 2 | Discrete uniform | [10, 30] |
| | Group 2 proportion | perm grp 2 | Uniform | [0.7, 1.0] |
| | Group 1 mean | perm mean 1 | Uniform | [−1.5, 1.5] |
| | Group 2 mean | perm mean 2 | Uniform | [1.5, 3.0] |
| **Φ** | Horizontal range | poro range 1 | Discrete uniform | [40, 80] |
| | Vertical range | poro range 2 | Discrete uniform | [10, 30] |
| | Group 2 proportion | poro grp 2 | Uniform | [0.0, 0.3] |
| | Group 1 mean | poro mean 1 | Uniform | [0.15, 0.19] |
| | Group 2 mean | poro mean 2 | Uniform | [0.19, 0.23] |

**Table 1.** Large-scale geological and engineering parameters and prior distributions used for the adaptive stochastic sampling

**Figure 18** shows the overall misfit progress of the run, highlighting the 5 best iterations of the run, along with their misfits. As a way of visualizing the match improvements obtained with the proposed method, the misfit calculated from the reference scenario is also shown.

The misfit evolutions for field production data (**Fig. 18**), shows that convergence occurred at around iteration 100, when four of the five lowest misfits were obtained. The lowest misfit belongs to iteration 91, with a misfit value of 32.01. Notice the significant improvement when compared with the reference scenario misfit value (388) and the lowest misfit obtained by conventional GHM using the fault–streamline regionalization (44) as shown in the previous section.

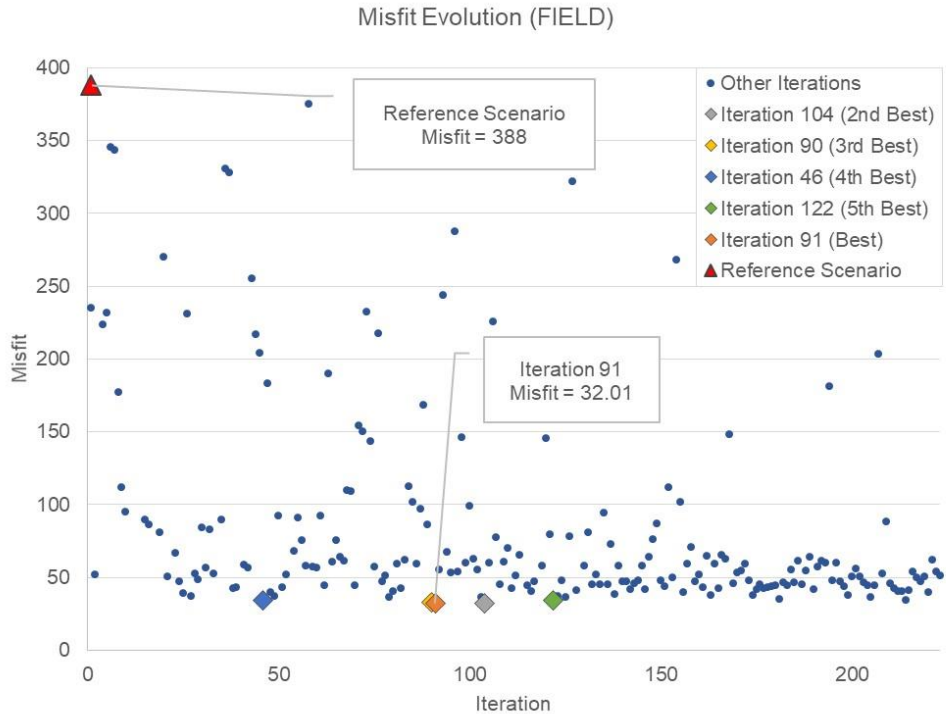

**Fig. 18.** Misfit evolution of the regionalization–based GHM algorithm, coupled with adaptive stochastic sampling

The fluid flow response for the best 5 iterations of the run in terms of field production data (**Fig. 19**) shows similar responses able to reproduce the observed production data particularly well for field oil and gas production rates while struggling to reproduce the final time steps of water production.

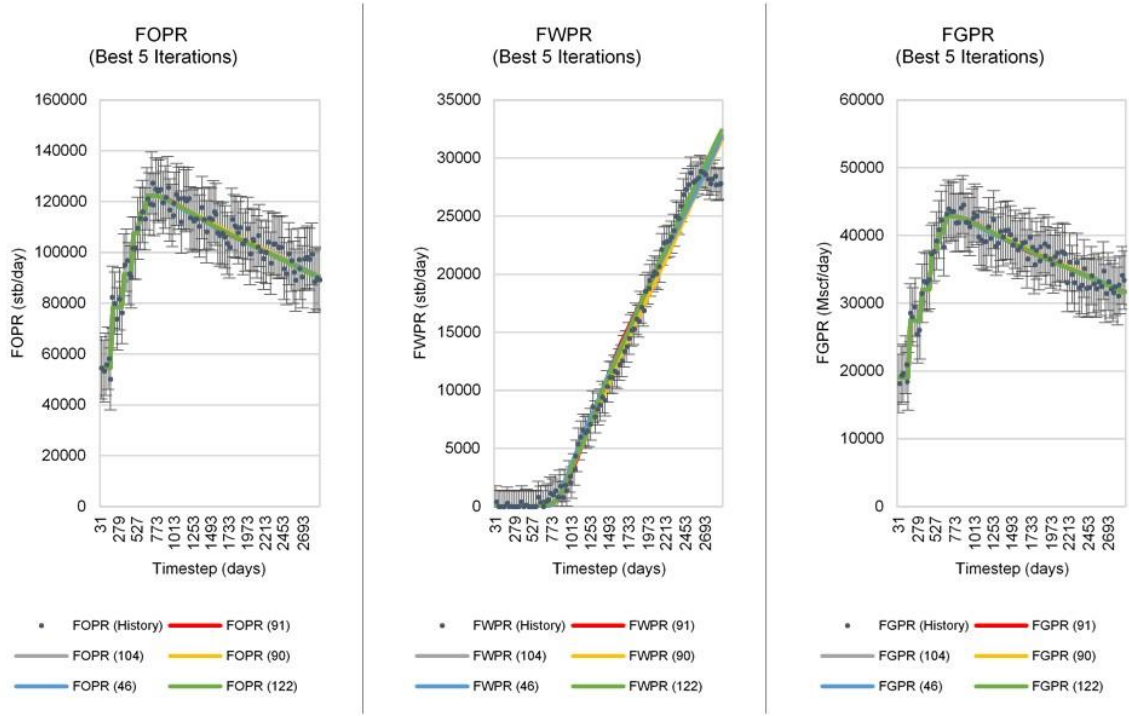

**Fig. 19.** Fluid flow response for FOPR (left), FWPR (middle) and FGPR (right) for GHM coupled with adaptive stochastic sampling

Due to the regionalization nature of the proposed approach it is important to assess the misfit evolution per region (**Fig. 20**) as it provides insight on the relationship between local production mismatches and local geological and engineering property perturbation. Regions 5, 7 and 8 show a convergence pattern along the iterative procedure, meaning that the generated models are locally adequate in terms of matching production. On the other hand, the remaining regions show erratic or divergent behavior. These are unmatched regions locally and therefore associated with higher uncertainty. This contradictory behavior can be explained by the fact that some regions have more than one well associated and by water production shifting from one well to another (i.e., from region-to-region) depending on changes on the spatial distribution of the petrophysical models generated at a given iteration. Looking at the misfit values for Region 2, 5 and 7 (**Fig. 20**), it is possible to observe that misfit values start dropping at around iteration 30. As a consequence, there is a shift of water production to a different region. This effect is noticed for example for Region 3 between iterations 30 to 50. When misfit values for Region 3 decrease at around iteration 50, a sudden increase in misfit is observed in Region 6. The misfit pattern per region illustrates the shifting of water production, occurring in early iterations, from Region 2, 5 and 7 to Region 3 and then to Region 6. The final pattern shows Region 6 as producing the excessive water in the reservoir, which resulted preferential in terms of achieving lower global misfit values throughout the remainder of the run.

On the contrary, some regions do not converge (Regions 1, 4 and 10; **Fig. 20**), meaning that the conditional assimilation of petrophysical properties and the adopted parameter perturbation from adaptive stochastic sampling is missing the description of the underlying geology to allow the match on these regions. At the same time, using single objective optimization for PSO may not fully capture local misfit convergence. Better results could be achieved by a using multi objective optimization approach in order to balance the local mismatch between multiple regions (Hutahaean et al. 2017).

**Figure 20** also shows that the proposed methodology is able to explore the parameter search space in a non-exploitative way, finding multiple parameter combinations for equally good misfits and defocusing on specific regional convergence to account for better global match alternatives.

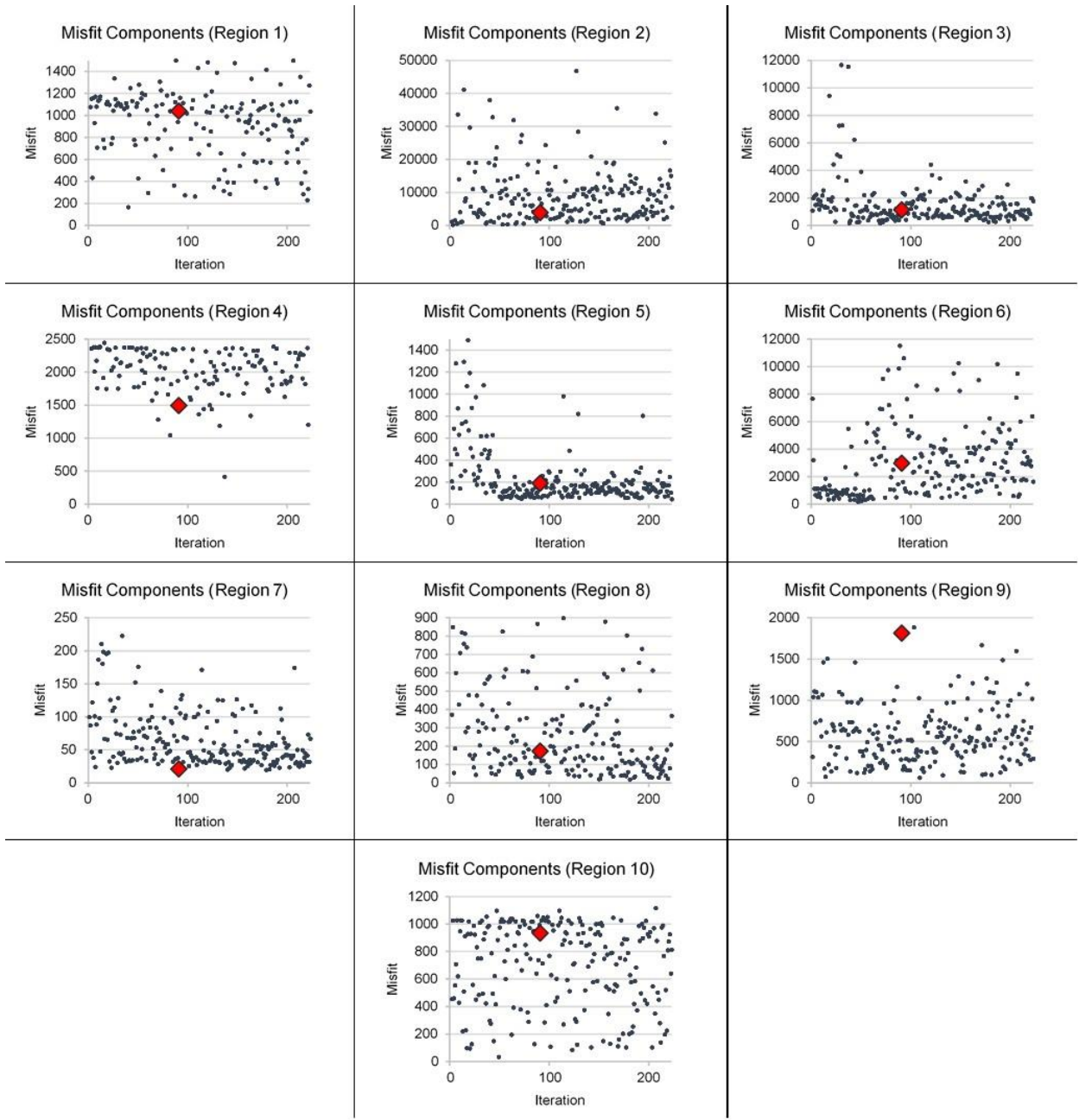

**Fig. 20.** Misfit evolution over different regions (**Fig. 10**) (best iteration is shown in red diamond)

**Figure 21** shows an example of the 5 best matches for oil and water production rates for WELL5 and WELL7 illustrating production match quality, with flow rates being matched throughout the total production time.

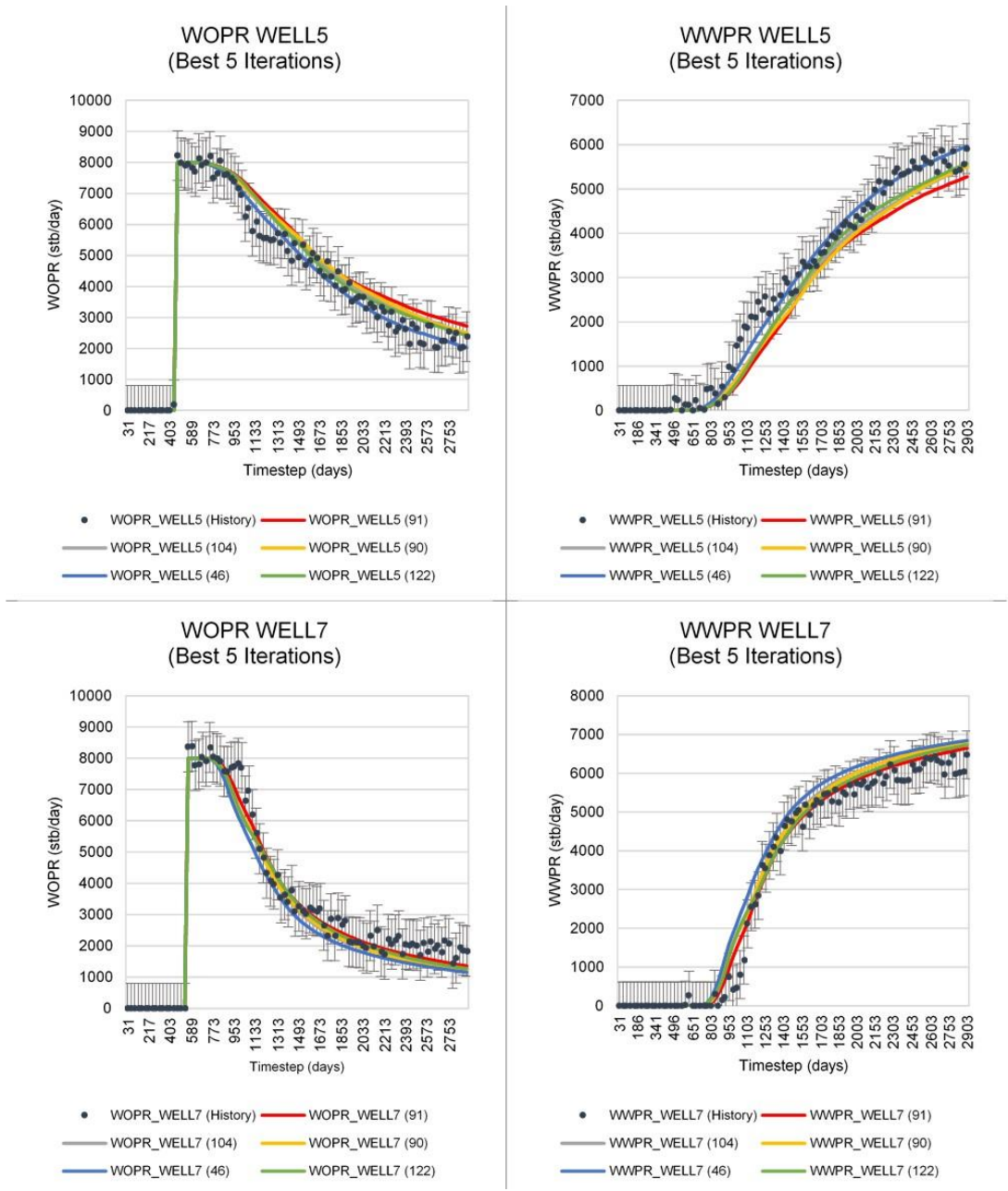

**Fig. 21.** Oil and water production rates for WELL5 (top left, top right) and WELL7 (bottom left, bottom right) for the 5 best iterations

Besides the misfit evolution, it is important to assess how each parameter evolves during the iterative procedure. Fault transmissibilities (**Fig. 22**) show a large variability within the ensemble of simulated values, showing a sinuous pattern during the model parameter space search for faults

2, 8 and 9. This can be explained by changes in the spatial distribution pattern of the static model whenever a best misfit value is achieved along the course of the iterative procedure. These changes correspond to a significant deviation from previous iterations, in terms of dynamic response. This may sometimes force the PSO sampling algorithm to search the parameter space for a combination of parameter values that will go into a different direction from what was previously being observed. Alternatively, this behavior may indicate that these particular faults may not be significant enough regarding the dynamic response of the model. On the contrary, fault 3, 5, 6 and 7 show a convergence towards a smaller range of possible values.

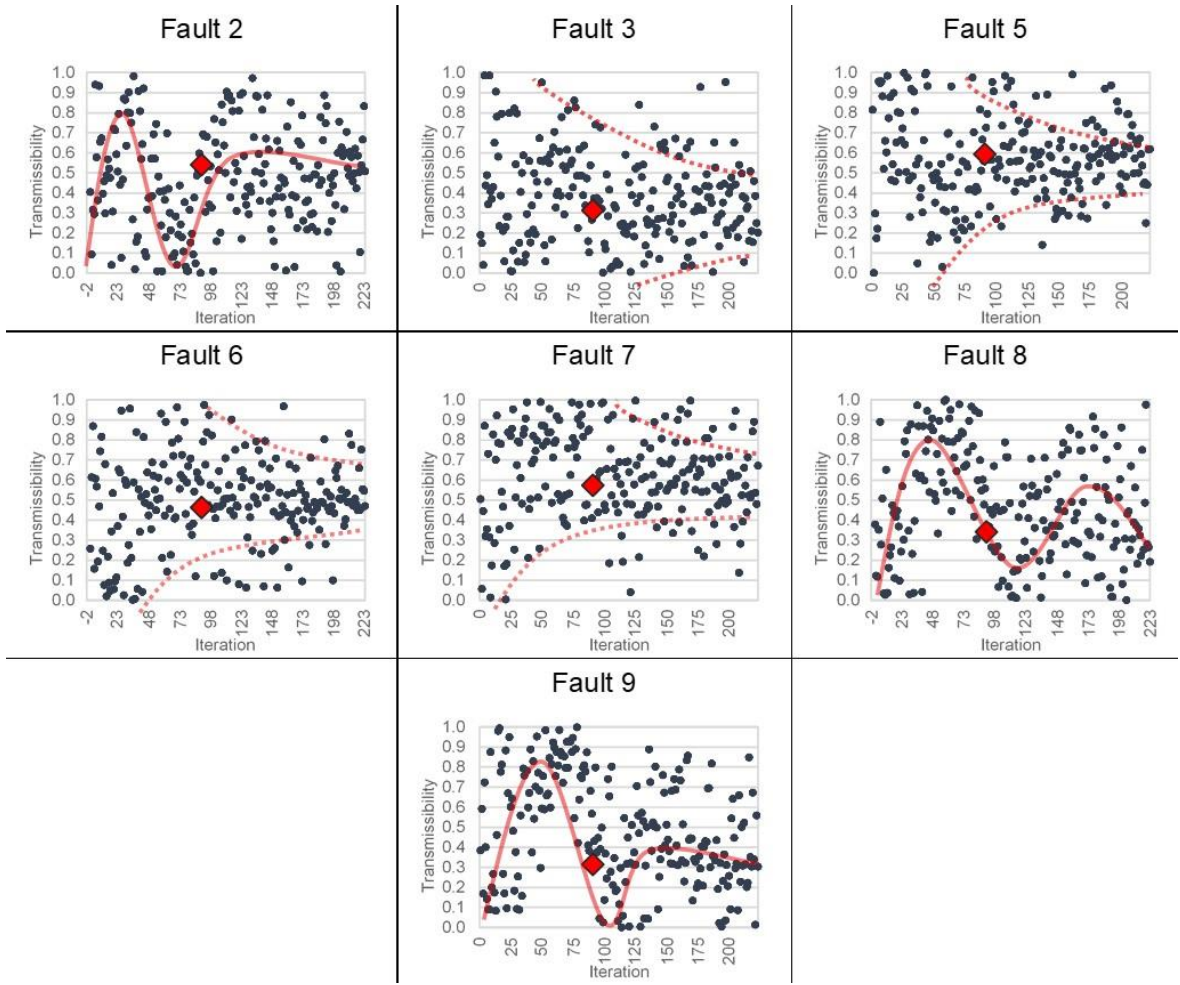

**Fig. 22.** Fault transmissibility sampling over iteration (best iteration is shown in red diamond)

**Figure 23** shows the parameter value versus misfit score in what fault transmissibility concerns. It shows that considerably different values reach similar misfit scores, allowing to infer that this is highly uncertain parameter. However, for faults 3, 5 and 6 it is possible to interpret a convergence towards a smaller range of values associated with a lower misfit. In these cases, the uncertainty decreases from the prior distribution into specific ranges. For fault 6 (**Fig. 4**), since it is located across the main producing region of the reservoir (separating the majority of producer wells), the PSO algorithm converged to an average value of 0.5, in effect, converging to the best balance between transmissibility value and the different production responses obtained during the course of the run. In an opposite way, the optimization for fault 2 transmissibility (**Fig. 4**), which mainly controls region 3 and region 6 (**Fig. 10**), shows a parameter sampling pattern that goes in hand with the shift in the misfit of those regions. At around iteration 25, when high transmissibility values are sampled for fault 2, the misfit for region 3 suddenly increases, consisting in the highest change in the misfit for all regions. The PSO algorithm detects this change and corrects its sampling choice to lower values of transmissibility, this time causing the misfit for region 6 to increase, again the most dramatic change in the misfit occurring at that stage of the run. The PSO algorithm reacts again to this by sampling for transmissibility values from higher ranges of the prior distribution and afterwards by converging again with less strength to lower values.

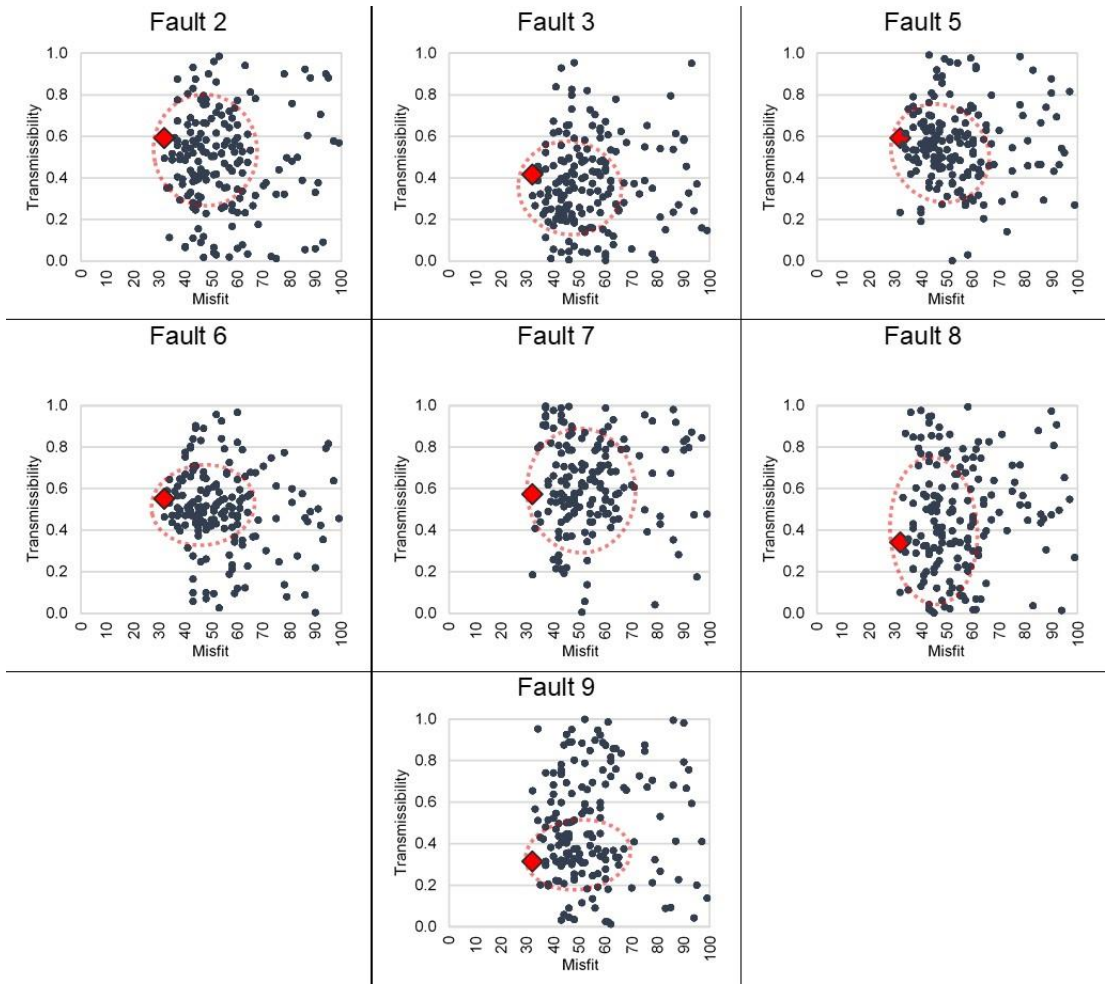

**Fig. 23.** Fault transmissibility parameter value versus misfit (best iteration is shown in red diamond)

Concerning the horizontal and vertical ranges of the variogram models for K (**Fig. 24**), it is possible to observe a decreasing trend from iteration to iteration towards the lower bound of the prior distribution assumed for this parameter. It is likely that the lower bound defined for this parameter is too restrictive, indicating that a smaller value would allow the convergence to lower values for horizontal K variogram range parameters. In terms of K vertical ranges (**Fig. 24**), its convergence is not so noticeable, nevertheless, a trend for sampling values in the center of the a priori distribution can be interpreted on the final iterations. From the interpretation of these results it can also be concluded that, fluid flow for this case study occurs predominantly along the horizontal plane defined by the variogram ranges and the resulting petrophysical spatial continuity of the realizations along that plane, which consequently shows that there is little sensitivity to K variogram ranges along the vertical direction. This is consistent with the reservoir geometry and with well location, which occurs predominantly along the horizontal plane.

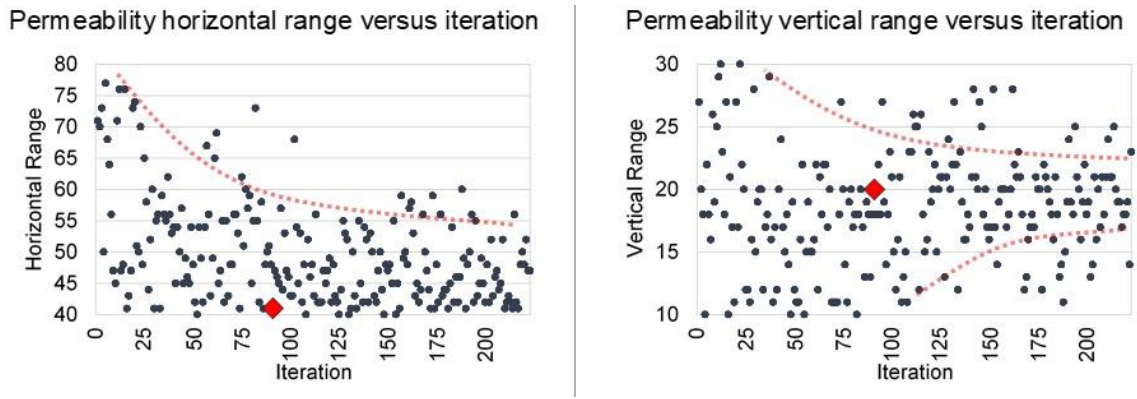

**Fig. 24.** K range versus iteration (left – horizontal range, right – vertical range, best iteration is shown in red diamond)

The same behavior can be interpreted from the parameter value versus misfit score (**Fig. 25**). There is a clear convergence of low misfit scores towards small horizontal variogram ranges. For the vertical variogram range, the spread of parameter values corresponding to lower misfits is higher, resulting in a higher uncertainty on this parameter.

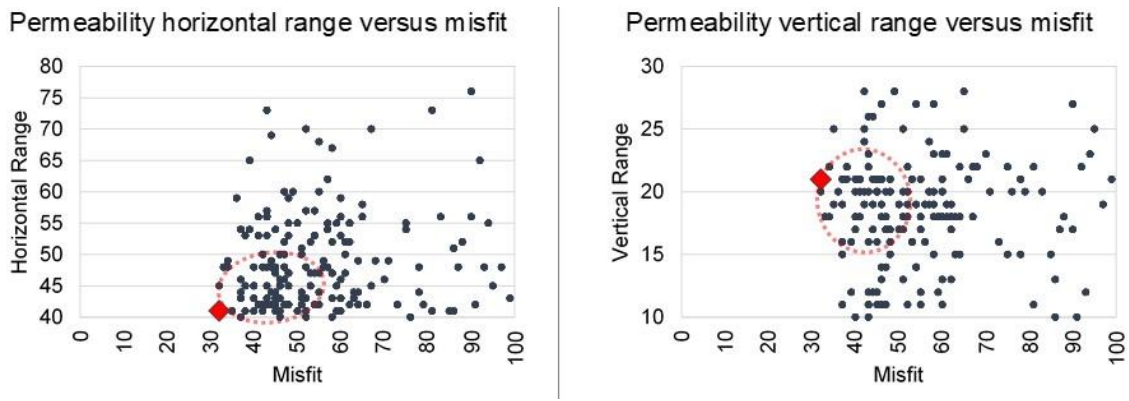

**Fig. 25.** K range versus misfit (left – horizontal range, right – vertical range, best iteration is shown in red diamond)

For the parameters related to the distribution of K (**Fig. 26**) there is a trend for sampling higher values for proportions of group 2, which is the property group associated with the highest K values (Sect. 3.1). Regarding the mean values for both property groups, it is possible to observe that there is not a clear trend for group 1. For group 2, convergence is observed towards a range of values located at the mid to top range of the perturbation distribution interval.

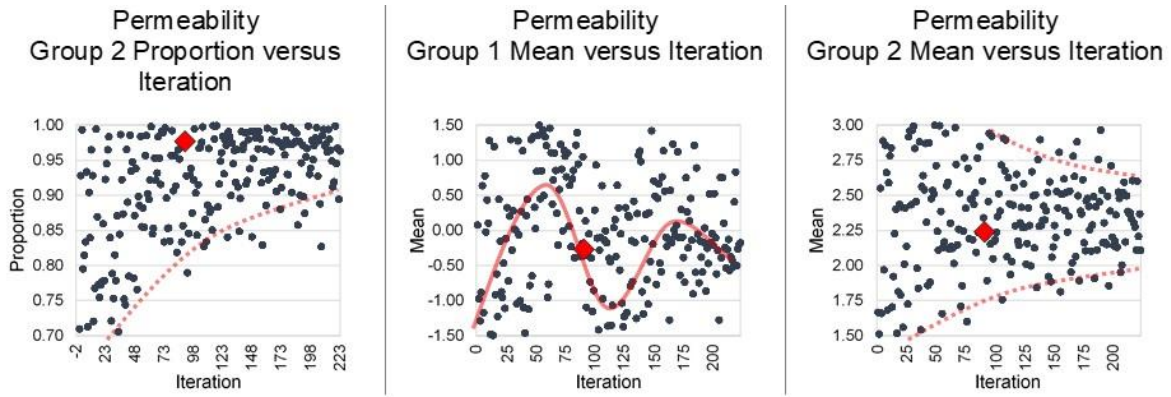

**Fig. 26.** K Histogram perturbation versus iteration, (left– proportion of property group 2, middle – mean for property group 1, right – mean for property group 2, best iteration is shown in red diamond)

**Figure 27** shows the parameters related to K distribution against the misfit score. There is a convergence of values for the proportion group 2 occurring at higher values. The mean K for group 1 does not have an identifiable pattern, meaning that the perturbation for this parameter does not have an effect that is relevant enough on dynamic response. This can be confirmed by the histogram of the reference K model that has a single mode. As for the mean of group 2, it is possible to interpret convergence at the top half part of the parameter distribution interval.

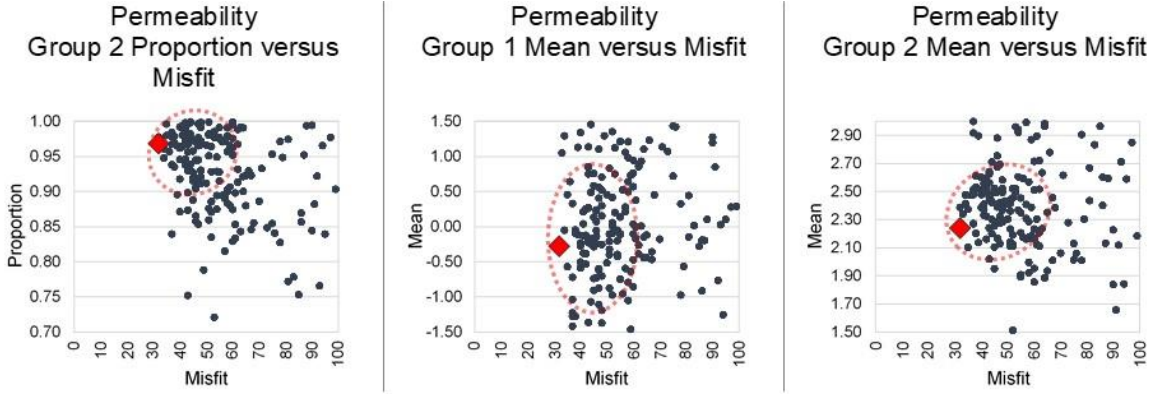

**Fig. 27.** K histogram perturbation versus misfit, (left – proportion of property group 2, middle – mean for property group 1, right – mean for property group 2, best iteration is shown in red diamond)

**Figure 28** compares the histograms as for $\log_{10}(K)$ as inferred form the scenario used as reference against the one retrieved from the best realization (iteration 91, simulation 5). Using exclusively stochastic sequential simulation as the perturbation engine of iterative geostatistical history matching (as in the previous section) does not allow extending the distribution limits, since these are reproduced exactly as retrieved from the available experimental data. The proposed approach allows the extension the limits of the distribution while keeping a single main population (i.e., unimodal distribution) at the same position.

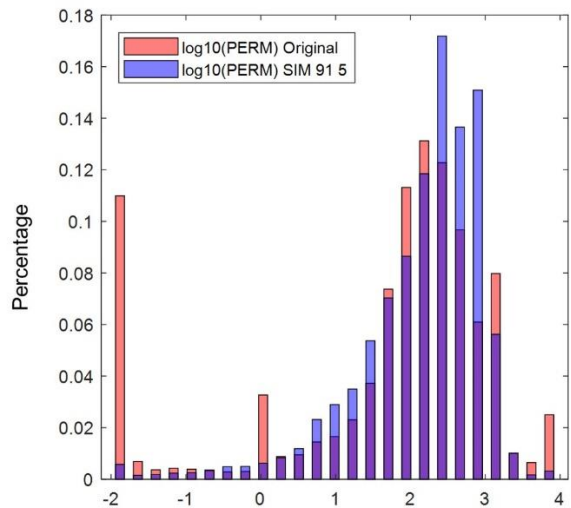

**Fig. 28.** Comparison between log10(K) histogram for reference scenario (red bars) and best iteration (blue bars)

The parameters perturbation related to the variogram model of Φ (**Fig. 29**) shows the values obtained for horizontal and vertical Φ ranges per iteration. A trend towards lower values of horizontal ranges can be observed, contrary to a trend toward higher values when considering the vertical range of the variogram model. This fact is supported when interpreting the values sampled during the iterative procedure against the misfit score (**Fig. 30**). A good estimate of the uncertainty related to this parameter may be inferred by the range of values obtained for smaller misfit values.

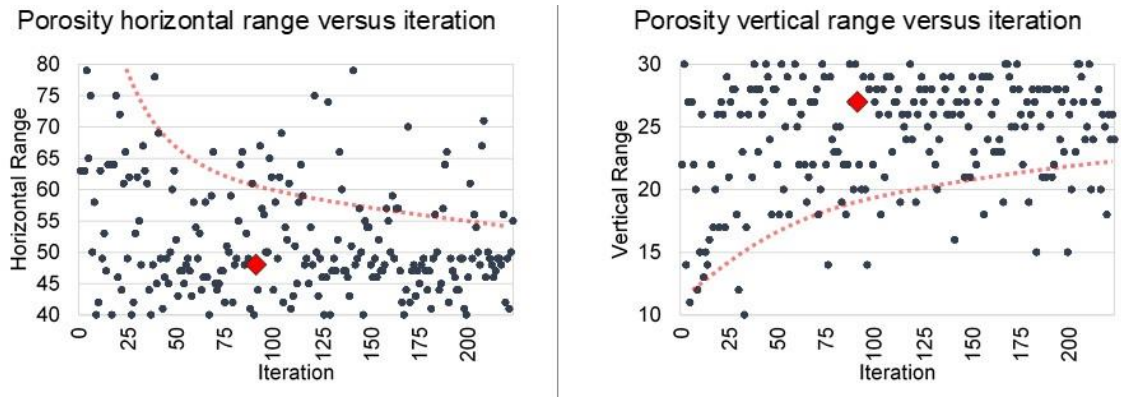

**Fig. 29.** Φ range versus iteration (left – horizontal range, right – vertical range)

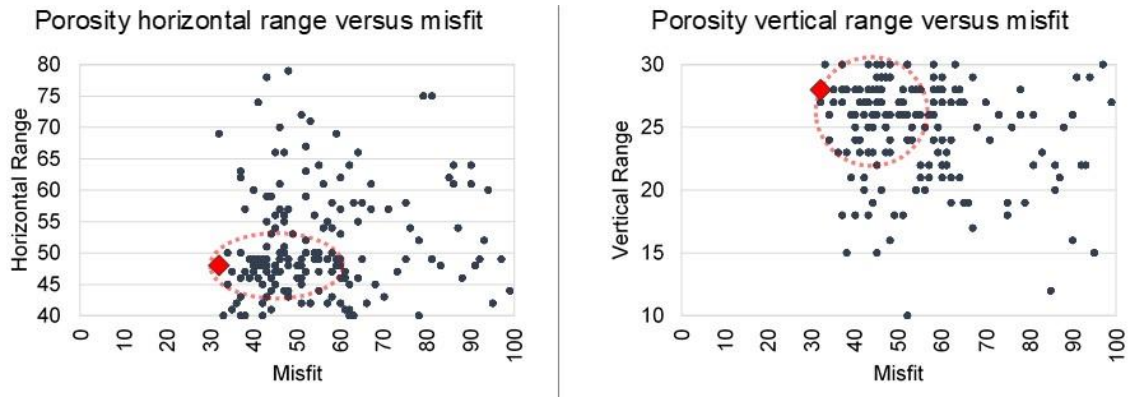

**Fig. 30.** Φ range versus misfit (left – horizontal range, right – vertical range)

Regarding the perturbation of the parameters of Φ related to its distribution (**Fig. 31**), both the proportions and property group means tend to converge into a smaller range of plausible values when compared with the prior distributions. The misfit score versus the parameter values (**Fig. 32**) shows that from the three parameters, the mean of the distribution associated with group 2 tends to be the preponderant variable. The final convergence obtained showed a wider range (between 0.1 and 0.2) for Group 2 proportions, with Group 1 consisting of high end models (over 0.17 Φ) and Group 2 models having a narrow range of Φ values (from 0.21 to 0.22).

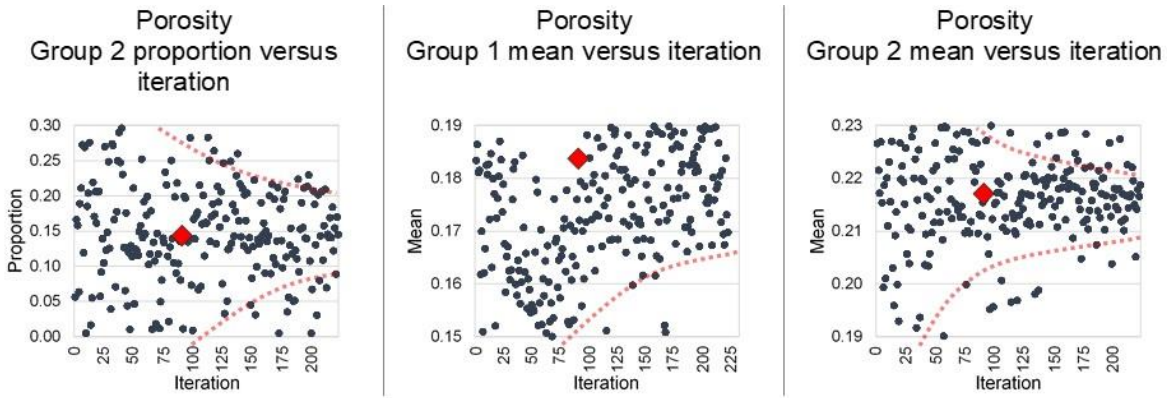

**Fig. 31.** Φ histogram perturbation versus iteration, (left – proportion of property group 2, middle – mean for property group 1, right – mean for property group 2, best iteration is shown in red diamond)

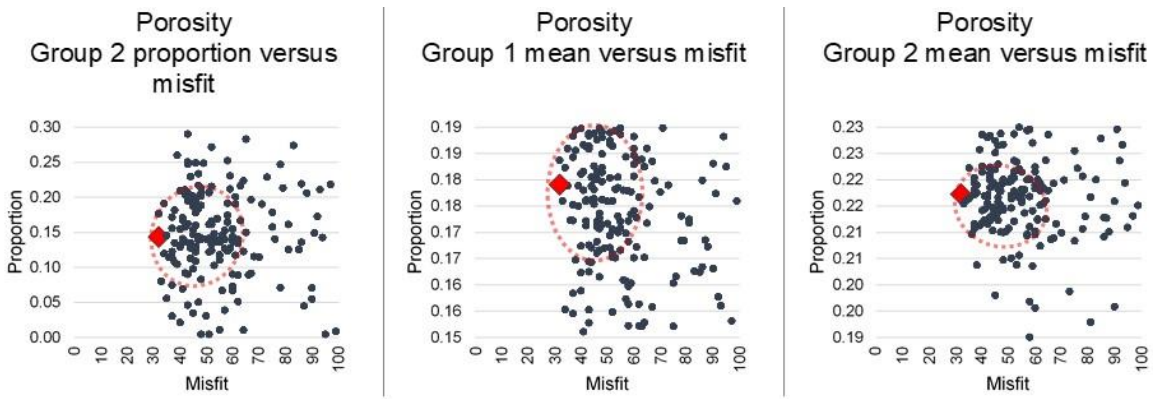

**Fig. 32.** Φ histogram perturbation versus misfit, (left – proportion of property group 2, middle – mean for property group 1, right – mean for property group 2, best iteration is shown in red diamond)

The histogram resulting from the best realization (**Fig. 33**) capture the main mode of the reference histogram without the need to extend the existing minimum and maximum.

Finally, horizontal sections extracted for both properties from the best-fit model (**Fig. 34** and **Fig. 35**), show a considerably different spatial continuity pattern when compared with those retrieved by the conventional geostatistical history matching being able to generate more geologically consistent model. The large-scale features of interest also resemble the reference model for both properties. It is important to highlight that this reference model is not the true model but a single possibility among a set of geological scenarios available in the original Watt field dataset. The proposed methodology was able to generate better matches by tuning uncertain large-scale geological and engineering parameters in a consistent regionalization domain. By doing this, the assumption of stationarity intrinsic of the stochastic sequential simulation algorithms that limit the exploration of the model parameter space can be avoided, resulting in a bias of the set of models retrieved at the end of the history matching procedure.

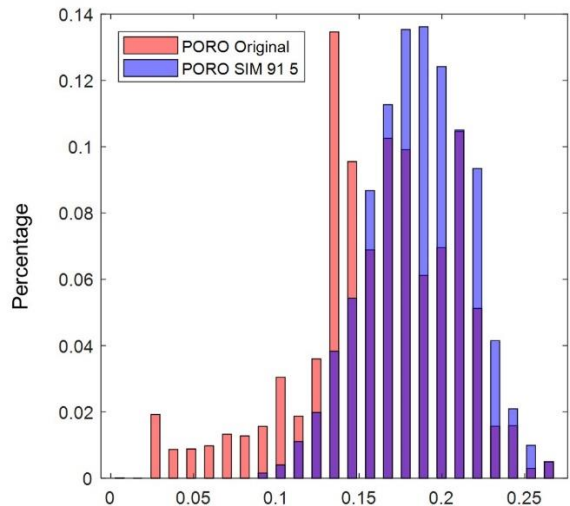

**Fig. 33.** Comparison between $\Phi$ histogram for base scenario (red bars) and best iteration (blue bars)

Finally, horizontal sections extracted from the best K field (**Fig. 34**) and Φ field (**Fig. 35**) are shown.

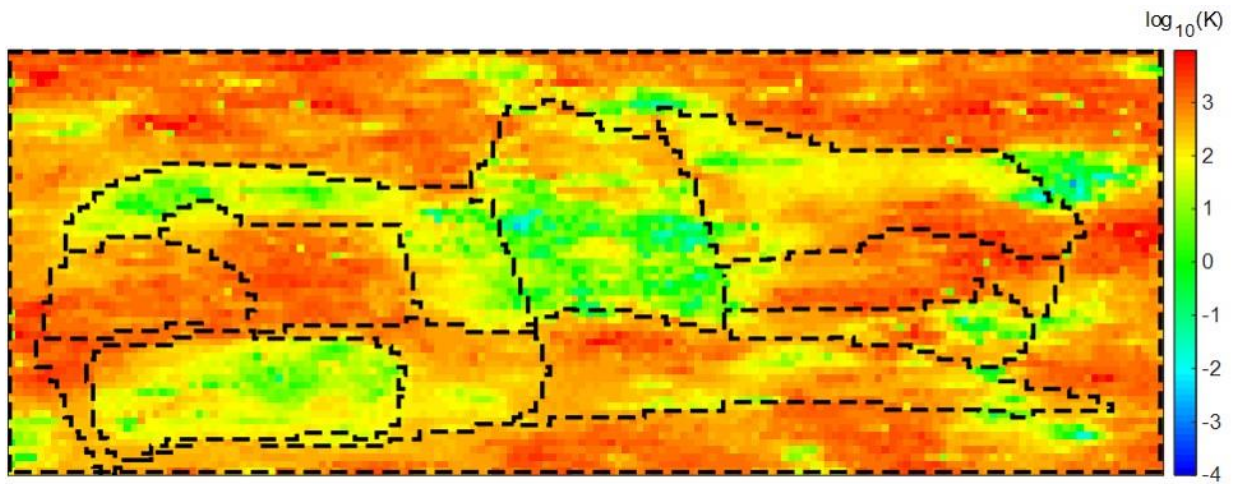

**Fig. 34.** Top view of the best-fit K realization (iteration 91, simulation 5)

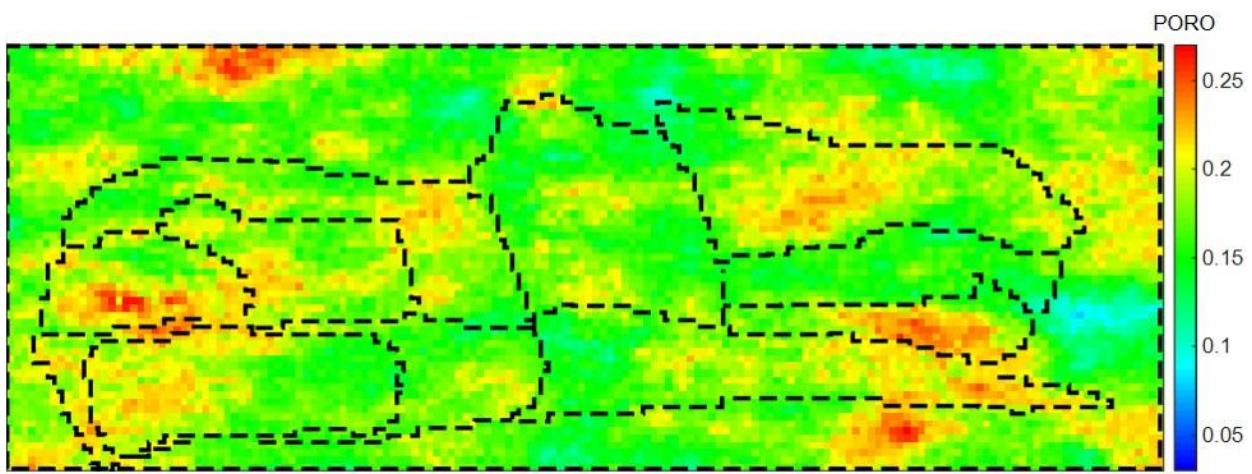

**Fig. 35.** Top view of the best-fit Φ realization (iteration 91, simulation 5)

## 4 Conclusions

This paper introduces a geological and dynamic consistent regionalization approach for geostatistical history matching. History matching is achieved through geostatistical assimilation of perturbed local regions between multiple stochastic realizations. The comparative analysis demonstrated that a more geologically consistent geometry of the perturbed regions provides better history match as opposed to a pure geometric (rectangular, circular, Voronoi) regionalization. The proposed geostatistical history matching approach allows to infer geological uncertainty both at local and global scales. The latter accounts for uncertainty in spatial correlation ranges, orientation and the a priori global property distribution (porosity and permeability). These large-scale geological parameters are perturbed with adaptive stochastic sampling - particle swarm optimization in geostatistical history matching loops coupled with assimilation of the local perturbation between sequential geostatistical realizations of porosity and permeability distributions. The results show the ability of the proposed technique in inferring alternative geological scenarios representing the uncertainty associated with given geological and dynamic parameters (e.g., some fault transmissibility values are irrelevant for the matching parameters) without compromising the match between simulated and observed production curves.

**Acknowledgements**

The authors gratefully acknowledge the support of the CERENA (strategic project FCT-UID/ECI/04028/2013). Erasmus+ for supporting the MSc traineeship with Heriot-Watt University, Rock Flow Dynamics for providing the licenses for tNavigator as well as MathWorks and Schlumberger for the donation of academic licenses for, respectively, Petrel® and MATLAB®. The Watt field data for this study was collated by Dan Arnold and provided on behalf of Heriot-Watt University.

**References**

Arnold, D., Demyanov, V., Tatum, D., Christie, M., Rojas, T., Geiger, S., & Corbett, P. (2013). Hierarchical benchmark case study for history matching, uncertainty quantification and reservoir characterisation. *Computers & Geosciences*, *50*, 4–15.

Ballester, P. J., & Carter, J. N. (2007). A parallel real-coded genetic algorithm for history matching and its application to a real petroleum reservoir. *Journal of Petroleum Science and Engineering*, *59*(3–4), 157–168. https://doi.org/10.1016/j.petrol.2007.03.012

Barrela, E., Azevedo, L., Soares, A., & Guerreiro, L. (2018). Integrating consistent streamline regionalization and multi-local distribution functions into geostatistical history matching: A real case application. In *RDPETRO 2018: Research and Development Petroleum Conference and Exhibition* (pp. 136–139). Abu Dhabi, UAE. https://doi.org/10.1190/RDP2018-35367099.1

Caeiro, M. H., Demyanov, V., & Soares, A. (2015). Optimized History Matching with Direct Sequential Image Transforming for Non-Stationary Reservoirs. *Mathematical Geosciences*, *47*(8), 975–994. https://doi.org/10.1007/s11004-015-9591-0

Caers, J., & Hoffman, T. (2006). The Probability Perturbation Method: A New Look at Bayesian Inverse Modeling. *Mathematical Geology*, *38*(1), 81–100. https://doi.org/10.1007/s11004-005-9005-9

Carneiro, J., Azevedo, L., & Pereira, M. (2018). High-dimensional geostatistical history matching: Vectorial multi-objective geostatistical history matching of oil reservoirs and uncertainty assessment. *Computational Geosciences*, *22*(2), 607–622. https://doi.org/10.1007/s10596-017-9712-6

Carrera, J., Alcolea, A., Medina, A., Hidalgo, J., & Slooten, L. J. (2005). Inverse problem in hydrogeology. *Hydrogeology Journal*, *13*(1), 206–222. https://doi.org/10.1007/s10040-004-0404-7

Cosentino, L. (2001). *Integrated reservoir studies*. Paris: Editions Technip.

de Sousa, S. H. G. (2007). Scatter Search Metaheuristic Applied to the History Matching Problem. Society of Petroleum Engineers. https://doi.org/10.2118/113610-STU

Demyanov, V., Backhouse, L., & Christie, M. (2015). Geological feature selection in reservoir modelling and history matching with Multiple Kernel Learning. *Computers & Geosciences*, *85*, 16–25. https://doi.org/10.1016/j.cageo.2015.07.014

Demyanov, V., Pozdnoukhov, A., Kanevski, M., & Christie, M. (2008). Geomodelling of a fluvial system with semi-supervised support vector regression. In *Proceedings of the VII international geostatistics congress, GECAMIN, Chile* (pp. 627–636).

Demyanov, Vasily, Arnold, D., Rojas, T., & Christie, M. (2018). Uncertainty Quantification in Reservoir Prediction: Part 2—Handling Uncertainty in the Geological Scenario. *Mathematical Geosciences*. https://doi.org/10.1007/s11004-018-9755-9

Demyanov, Vasily, Christie, M., Kanevski, M., & Pozdnoukhov, A. (2012). Reservoir Modelling Under Uncertainty-A Kernel Learning Approach.

Duane, S., Kennedy, A. D., Pendleton, B. J., & Roweth, D. (1987). Hybrid Monte Carlo. *Physics Letters B*, *195*(2), 216–222. https://doi.org/10.1016/0370-2693(87)91197-X

Emerick, A. A., & Reynolds, A. C. (2013). Ensemble smoother with multiple data assimilation. *Computers & Geosciences*, *55*, 3–15. https://doi.org/10.1016/j.cageo.2012.03.011

Erbas, D., & Christie, M. A. (2007). Effect of Sampling Strategies on Prediction Uncertainty Estimation. Society of Petroleum Engineers. https://doi.org/10.2118/106229-MS

Evensen, G., Hove, J., Meisingset, H., Reiso, E., Seim, K. S., & Espelid, Ø. (2007). Using the EnKF for Assisted History Matching of a North Sea Reservoir Model. Society of Petroleum Engineers. https://doi.org/10.2118/106184-MS

Hajizadeh, Y., Christie, M. A., & Demyanov, V. (2011). Towards Multiobjective History Matching: Faster Convergence and Uncertainty Quantification. Society of Petroleum Engineers. https://doi.org/10.2118/141111-MS

Hoffman, B. T., & Caers, J. (2003). Geostatistical History Matching Using a Regional Probability Perturbation Method. Society of Petroleum Engineers. https://doi.org/10.2118/84409-MS

Horta, A., & Soares, A. (2010). Direct Sequential Co-simulation with Joint Probability Distributions. *Mathematical Geosciences*, *42*(3), 269–292. https://doi.org/10.1007/s11004-010-9265-x

Hu, L. Y., Blanc, G., & Noetinger, B. (2001). Gradual Deformation and Iterative Calibration of Sequential Stochastic Simulations. *Mathematical Geology*, *33*(4), 475–489. https://doi.org/10.1023/A:1011088913233

Hutahaean, J., Demyanov, V., & Christie, M. A. (2017). On Optimal Selection of Objective Grouping for Multiobjective History Matching. *SPE Journal*, *22*(04), 1296–1312. https://doi.org/10.2118/185957-PA

Kathrada, M. (2009). *Uncertainty evaluation of reservoir simulation models using particle swarms and hierarchical clustering*. Heriot-Watt University.

Kazemi, A., & Stephen, K. D. (2013). Optimal Parameter Updating in Assisted History Matching Using Streamlines as a Guide. *Oil & Gas Science and Technology – Revue d'IFP Energies Nouvelles*, *68*(3), 577–594. https://doi.org/10.2516/ogst/2012071

Kennedy, J., & Eberhart, R. (1995). Particle swarm optimization (Vol. 4, pp. 1942–1948). IEEE. https://doi.org/10.1109/ICNN.1995.488968

Le Ravalec-Dupin, M., & Da Veiga, S. (2011). Cosimulation as a perturbation method for calibrating porosity and permeability fields to dynamic data. *Computers & Geosciences*, *37*(9), 1400–1412. https://doi.org/10.1016/j.cageo.2010.10.013

Mata-Lima, H. (2008). Reducing uncertainty in reservoir modelling through efficient history matching. *Oil Gas European Magazine*, *3*, 2008.

Mohamed, L., Christie, M. A., & Demyanov, V. (2009). Comparison of Stochastic Sampling Algorithms for Uncertainty Quantification. Society of Petroleum Engineers. https://doi.org/10.2118/119139-MS

Nunes, R., Soares, A., Azevedo, L., & Pereira, P. (2016). Geostatistical Seismic Inversion with Direct Sequential Simulation and Co-simulation with Multi-local Distribution Functions. *Mathematical Geosciences*, *49*(5), 583–601. https://doi.org/10.1007/s11004-016-9651-0

Oliver, D. S., & Chen, Y. (2011). Recent progress on reservoir history matching: a review. *Computational Geosciences*, *15*(1), 185–221. https://doi.org/10.1007/s10596-010-9194-2

Oliver, D. S., Reynolds, A. C., & Liu, N. (2008). *Inverse Theory for Petroleum Reservoir Characterization and History Matching*. Leiden: Cambridge University Press. Retrieved from http://public.eblib.com/choice/publicfullrecord.aspx?p=343536

Roggero, F., & Hu, L. Y. (1998). Gradual Deformation of Continuous Geostatistical Models for History Matching. Society of Petroleum Engineers. https://doi.org/10.2118/49004-MS

Sambridge, M. (1999). Geophysical inversion with a neighbourhood algorithm-I. Searching a parameter space. *Geophysical Journal International*, *138*(2), 479–494. https://doi.org/10.1046/j.1365-246X.1999.00876.x

Schulze-Riegert, R. W., Axmann, J. K., Haase, O., Rian, D. T., & You, Y.-L. (2001). Optimization Methods for History Matching of Complex Reservoirs. Society of Petroleum Engineers. https://doi.org/10.2118/66393-MS

Vargas-Guzmán, J. A., Al-Gaoud, A., Datta-Gupta, A., Jiménez, E. A., & Oyeriende, D. (2009). Identification of high permeability zones from dynamic data using streamline simulation and inverse modeling of geology. *Journal of Petroleum Science and Engineering*, *69*(3–4), 283–291. https://doi.org/10.1016/j.petrol.2009.09.004

Vincent, G., Corre, B., & Thore, P. (1999). Managing structural uncertainty in a mature field for optimal well placement. *SPE Reservoir Evaluation & Engineering*, *2*(04), 377–384.